\documentclass[11pt,a4paper]{article}
\usepackage{jheppub}

\newcommand{\lbl}[1]{\label{eq:#1}}
\newcommand{ \rf}[1]{(\ref{eq:#1})}

\newcommand{\be}{\begin{equation}}
\newcommand{\ee}{\end{equation}}
\newcommand{\bea}{\begin{eqnarray}}
\newcommand{\eea}{\end{eqnarray}}
\newcommand{\setl}{\setlength\arraycolsep{2pt}} 

\newcommand{\noi}{\noindent}
\newcommand{\nn}{\nonumber}
\newcommand{\ra}{\rightarrow}

\newcommand{\cA}{{\cal A}}
\newcommand{\cB}{{\cal B}}
\newcommand{\cC}{{\cal C}}

\newcommand{\cM}{{\cal M}}

\newcommand{\cO}{{\cal O}}

\newcommand{\cR}{{\cal R}}

\newcommand{\cE}{{\cal E}}

\newcommand{\Imm}{\mbox{\rm Im}}
\newcommand{\Ree}{\mbox{\rm Re}}
\newcommand{\annd}{\mbox{\rm and}}
\newcommand{\foor}{\mbox{\rm for}}
\newcommand{\MeV}{\mbox{\rm MeV}}
\newcommand{\GeV}{\mbox{\rm GeV}}

\title{Hadronic Vacuum Polarization and the MUonE proposal}

\author[1]{David Greynat}
\author[2]{Eduardo de Rafael}

\affiliation[1]{No affiliation at present}
\affiliation[2]{Aix-Marseille Univ, Universit\'{e} de Toulon, CNRS, CPT, Marseille, France}

\emailAdd{david.greynat@gmail.com}
\emailAdd{EdeR@cpt.univ-mrs.fr}			

\abstract{The MUonE proposal  at the CERN SPS consists in extracting the value of the hadronic vacuum polarization self-energy  function (HVP)  from its contribution to the differential cross-section of elastic muon-electron scattering. The HVP contribution to the muon anomalous magnetic moment can then be obtained from a weighted integral of the measured HVP self-energy  function.  This,  however, requires  a knowledge of the HVP function in its full integration domain.  This paper discusses a procedure to reconstruct the HVP function in the regions not directly accessible to measurement. The method  is based on  the so-called transfer theorems, due to Flajolet and Odlyzko,  which we explain and adapt  to HVP.}

\begin{document}

\vspace*{2cm}

\makeatletter
\def\@fpheader{\relax}
\makeatother
\maketitle

\section{Introduction}

The measurements of the anomalous magnetic moment of the muon $a_{\mu}$,  made  at BNL ~\cite{E821} and more recently at Fermilab~\cite{FNAL,FL21}, give the results:
\be
a_{\mu}^{{\rm BNL}}=116~592~089(63)\times 10^{-11}\quad\annd\quad a_{\mu}^{{\rm FNAL}}=116~592~040(54)\times 10^{-11}\,.
\ee
They agree with each other at the level of  0.6 standard deviations ($0.6\sigma$) and their combined number
\be\lbl{eq:BNLFL}
a_{\mu}(2021)=116~592~061(41)\times 10^{-11}\,,
\ee
has the remarkable accuracy of 0.35 parts per million.

The theoretical evaluation of the same observable in the Standard Model   has been made to a comparable precision. 
The result
\be
a_{\mu}({\rm Th.WP})=116~591~810(43)\times 10^{-11}
\ee
is the {\it consensus theory number} reported in the 2020 White Paper (WP) of ref.~\cite{PhRe20}. When compared to the experimental number in Eq.~\rf{eq:BNLFL} it turns out to be  4.2$\sigma$ below, a significant difference, which has triggered many speculations in the literature~\cite{BSM21} on what kind of  new physics  could explain this difference.

The situation at present, however, is rather confusing. The same day that the results of the Fermilab muon g-2 collaboration were published, the journal Nature also published a new result of the Budapest-Marseille-Wuppertal (BMW) lattice QCD (LQCD) collaboration on the lowest order hadronic vacuum polarization contribution to the muon g-2. Their result~\cite{BMWmu} 
\be
a_{\mu}({\rm HVP})_{\rm BMW}=7~075(55)\times 10^{-11}\,,
\ee
differs from previous evaluations using  data-driven dispersion relations~\cite{Davier,Teubner}:
\be\lbl{eq:disperevs}
a_{\mu}({\rm HVP})_{\rm lowest~order}^{\rm ref.[{7}]} = 6~940(40)\times 10^{-11}\quad\annd\quad a_{\mu}({\rm HVP})_{\rm lowest~order}^{\rm ref.[{8}]}=6~928(24)\times 10^{-11}\,,
\ee
which are incorporated in the
 {\it consensus theory number} given above. The BMW-lattice QCD result reduces the total discrepancy with the experimental result in Eq.~\rf{eq:BNLFL} from 4.2$\sigma$ to 1.6$\sigma$. Still a discrepancy, but not significant  to argue evidence for new physics.
The result is under detailed examination and  one hopes to have  news on that in the near future. If the disagreement between LQCD and the experimental dispersive evaluations of the HVP persists, one will have to find the explanation for that. Because they involve integrals of different quantities, comparison of the two methods is difficult, although not impossible. 

In the meantime, the Fermilab Muon g-2 experiment expects to  reduce the error by a factor of four over the next four years, as more statistics accumulate. There is also a new experiment at the Japan Proton Accelerator Research Complex in Tokai, the J-PARC experiment E34~\cite{JPARC}, which will employ a new different technique to measure the muon anomaly. 

Another expected experiment is the  
 MUonE proposal  at the CERN SPS~\cite{CPTV,ABetal,MUonEP}. It   
 consists in extracting the value of the HVP self-energy  function in the Euclidean, from its contribution to the differential cross-section of elastic muon-electron scattering with muons at ${\rm E}_{\mu}=160~\GeV$ colliding on atomic electrons of a fixed low Z target~\cite{Betal}. The muon anomaly can then be obtained from a weighted integral of the measured HVP self-energy  function discussed below. The purpose of this paper is to present theoretical arguments concerning this interesting  proposal which should increase its potential impact.

The function we shall be concerned with is the Fourier transform of the vacuum expectation value of the time-ordered product of two electromagnetic hadronic  currents of the Standard Model $J_{\mu}^{\rm had}(x)$ at two separate space-time $x$-points ($g_{\mu\nu}={\rm diag} (1,-1,-1,-1)$):

\be\lbl{eq:ft}
\Pi_{\mu\nu}^{\rm had}(q)=i\int_{-\infty}^{+\infty}\,  d^4 x\  e^{iq\cdot x}
\langle 0\vert T\left(J_{\mu}^{\rm had}(x)J_{\nu}^{\rm had}
(0)\right)\vert 0\rangle=(q_{\mu}q_{\nu}-q^2 g_{\mu\nu})\Pi_{\rm had}(q^2)\,.
\ee
The photon hadronic self-energy function $\Pi_{\rm had}(q^2)$  is a complex function of its  $q^2$ variable. It is an analytic function  in the full complex plane, but for a cut in the real axis which  goes  from the physical threshold at $q^2=4m_{\pi^\pm}^2\equiv t_0$ to infinity~\footnote{In the presence of higher order electromagnetic corrections the threshold is at the mass of the $\pi^0$ because of the $\pi^0 \gamma$ contribution to the spectral function. In this  paper the threshold is fixed at $t_0 =4m_{\pi^\pm}^2$, but can be adjusted to $m_{\pi^0}^2$ if necessary.} As such, the on-shell renormalized HVP-function,  i.e. $\Pi_{\rm had}(q^2)$ subtracted at its value at $q^2 =0$,  obeys the dispersion relation: 
\be\lbl{eq:Pi}
\Pi^{\rm HVP}(q^2)\equiv \Pi_{\rm had}(q^2)-\Pi_{\rm had}(0) =  \int_{t_0}^{\infty}\frac{dt}{t}\,
\frac{q^2}{t-q^2 -i\epsilon}\frac{1}{\pi}\Imm\Pi_{\rm had}(t)\,,\quad t_0 \equiv 4m_{\pi^\pm}^2\,. 
\ee

The optical theorem relates the hadronic spectral function $\frac{1}{\pi}\Imm\Pi_{\rm had}(t)$ to the observable one-photon annihilation cross-section:
\be
\sigma(t)_{e^+ e^- \ra {\rm had}} \underset{{m_e\ra 0}}{\thicksim}\frac{4\pi^2 \alpha}{t}\frac{1}{\pi}\Imm\Pi_{\rm had}(t)\,,
\ee
and this is the way that experimental data-driven determinations of $\frac{1}{\pi}\Imm\Pi_{\rm had}(t)$ have been obtained,  as well as the evaluation of  the anomalous magnetic moment of the muon $a_{\mu}^{\rm HVP}$, via the integral representation~\cite{BM61,BdeR,GdeR}
\be\lbl{eq:BMSR}
a_{\mu}^{\rm HVP}=\frac{\alpha}{\pi}\int_{t_0}^\infty \frac{dt}{t}\int_0^1 dx\  \frac{x^2 (1-x)}{x^2 +\frac{t}{m_{\mu}^2}(1-x)}\frac{1}{\pi}\ \Imm\Pi_{\rm had}(t)\,.
\ee

An alternative representation of  $a_{\mu}^{\rm HVP}$ in terms of the hadronic self-energy function $\Pi_{\rm had}(q^2)$ in the Euclidean ($Q^2\equiv -q^2 \ge 0$), proposed in refs.~\cite{LPdeR,EdeR94}, follows from a rearrangement of the integrand in Eq.~\rf{eq:BMSR} and the use of the dispersion relation in Eq.~\rf{eq:Pi}:

\bea
a_{\mu}^{\rm HVP} & = & \frac{\alpha}{\pi}\int_0^1 dx\  (1-x)\int_{t_0}^\infty \frac{dt}{t}\ \frac{\frac{x^2}{1-x}m_{\mu}^2}{t+\frac{x^2}{1-x}m_{\mu}^2}\ \frac{1}{\pi}\Imm\Pi_{\rm had}(t)\,, \nn \\ 
& = & -\frac{\alpha}{\pi}\int_0^1 dx\ (1-x)\ \Pi^{\rm HVP}\left( -\frac{x^2}{1-x}m_{\mu}^2\right)\,,\quad Q^2 \equiv \frac{x^2}{1-x}m_{\mu}^2\,.\lbl{eq:LdeR}
\eea

\noi
This $x$-Feynman parametric  representation   is particularly relevant to  the MUonE proposal~\footnote{It has also been proposed for   lattice QCD evaluations in ref.~\cite{Blum}.}.
It requires, however, the knowledge of the $x$-integrand in Eq.~\rf{eq:LdeR} in its full range and the experiment can only provide precise enough data in a limited $x$-window ($x_{\rm min}\simeq 0.3$ to $x_{\rm max}\simeq 0.9$). The obvious question which then arises is what reliable method can  be used  to extrapolate the determination in the $x$-window to the full $x$-integration domain. The purpose of this paper is to present a  systematic approximation procedure to achieve this reconstruction. It is   
based on what in the mathematical literature is known as  the transfer theorems of Flajolet and Odlyzko~\cite{FOth}~\footnote{For a comprehensive exposition  and related subjects see ref.~\cite{FS09}, in particular Part B. Complex Asymptotics.}. We shall refer to this as the procedure of {\it {\it reconstruction approximants}}.

\vspace*{0.5cm}

In the next section we review some of the HVP properties that we shall be using. The  transfer theorem adapted to HVP is discussed in section 3~\footnote{Another application of  transfer theorems,  within the context of heavy quarks in QCD, can be found  in refs.~\cite{GP,GMP,GM}.}. In section 4 we explain how to apply {\it reconstruction approximants}  in the case of the MUonE proposal, and in section 5 we illustrate this  with  a phenomenological model. Conclusions and outlook are given in Section 6.

Appendix A is dedicated to showing  how the {\it {\it reconstruction approximants}}  work in the case of the QED vacuum polarization at the one loop level, where all the steps can be followed  analytically. 
Appendix B contains   
technical details on  combinatorial analysis,  which have been used to derive some of the results in the text.

Readers who are only interested in the applications of {\it reconstruction approximants} are advised to  concentrate their attention on sections 4 and 5.

\section{Some Properties of HVP}

We shall often refer to the Mellin Transform of the hadronic spectral function
\begin{equation}
  \lbl{eq:defM}
  \mathcal{M}^{\rm HVP}(s) = \int_{t_0} ^\infty \frac{dt}{t} \left(\frac{t}{t_0}\right)^{s-1} \frac{1}{\pi} \Imm\Pi_\mathrm{had}(t)\,,\quad -\infty\le \Ree(s)<1\,,
\end{equation} 
with its domain of definition extended to the full complex $s$-plane by analytic continuation. The Mellin Transform $\mathcal{M}^{\rm HVP}(s)$ is then a meromorphic  function with poles in the real axis at  $\Ree(s)\ge 1$.

 Inserting the identity:
\be
\frac{1}{1+A}   = 
\frac{1}{2\pi i}\int\limits_{c_s-i\infty}^{c_s+i\infty}ds\ 
A^{-s}\ \Gamma(s)\Gamma(1-s)\quad {\rm with} \quad A\equiv\frac{Q^2}{t_0}\quad\annd\quad c_s \equiv \Ree(s) \in ]0,1[ \,,
\ee
in the integrand of the dispersion relation in  Eq.~\rf{eq:Pi},  
the HVP self-energy in the Euclidean ($Q^2 \ge 0$) can then  be expressed in terms of the 
 Mellin-Barnes integral~\footnote{Mellin-Barnes representations have been extensively discussed in ref.~\cite{FGD} and, within the context of HVP and $g_{ \mu}-2$, in ref.~\cite{ChGdeR} and references therein.  The precise definitions of   {\it fundamental strip} and {\it singular series}  can be found in ref.~\cite{FGD}.} 
\be\lbl{eq:inverseMB}
\Pi^{\rm HVP}(-Q^2) = -\frac{Q^2}{t_0}\ \frac{1}{2\pi i}\int\limits_{c_s-i\infty}^{c_s+i\infty}ds\ \left(\frac{Q^2}{t_0} \right)^{-s} \Gamma(s)\Gamma(1-s)\ \cM^{\rm HVP}(s)\,, \quad c_s \equiv \Ree(s) \in ]0,1[  \,.
\ee
The region $\Ree(s) \in ]0,1[$ in the complex $s$-plane  is called the {\it fundamental strip}, where the integral converges absolutely. 
Two basic properties of this representation  follow:

\begin{itemize}
\item

The Taylor expansion of $\Pi^{\rm HVP}(-Q^2)$ at small $Q^2$ is governed by the {\it singular series expansion}   of the integrand in Eq.~\rf{eq:inverseMB} at the {\it left} of the {\it fundamental strip}, i.e. $\Ree(s)\le 0$. In this case,  the {\it singular series} is   generated by the poles at $s=0,-1,-2,-3,\cdots$ of the $\Gamma(s)$ function in the integrand,  with the result: 
\begin{equation}
\lbl{eq:PiExpQ}
\Pi^{\rm HVP}(-Q^2) \underset{Q^2<t_0}{\sim} - \left(\frac{Q^2}{t_0}\right) \left[ \sum_{n=0}^\infty \left(\frac{Q^2}{t_0}\right)^{n} (-1)^n \; \mathcal{M}^{\rm HVP}(-n) \right]\;,
\end{equation}
where $\mathcal{M}^{\rm HVP}(-n)$ are the moments of the spectral function:
\be\lbl{eq:moments}
\mathcal{M}^{\rm HVP}(1-n)=\int_{t_0} ^\infty \frac{dt}{t} \left(\frac{t_0}{t}\right)^n \frac{1}{\pi} \Imm\Pi_\mathrm{had}(t)\,,\quad n=1,2,3,\cdots\ \,,
\ee
clearly accessible to experimental determination up to high-$t$ values, beyond which,  perturbative QCD (pQCD) can be used.

\item
By contrast, the asymptotic expansion of $\Pi^{\rm HVP}(-Q^2)$ at  large-$Q^2$ is governed by the {\it singular series expansion}   of the integrand in  Eq.~\rf{eq:inverseMB} at the {\it right} of the {\it fundamental strip}, i.e.  $\Ree(s)\ge 1$. This is a  series  of the form: 
\be\lbl{eq:sse}
\Gamma(s)\Gamma(1-s)\ \cM^{\rm HVP}(s)\asymp\ \sum_{\mathsf{p}=1,2,3,\dots}\ \sum_{k=0,1,2,\dots}\frac{	\mathsf{\cR}_{\mathsf{p},k}}{(s-\mathsf{p})^{k+1}}\,,
\ee
with $\mathsf{\cR}_{\mathsf{p},k}$  the residues of the singularities at $s=\mathsf{p}$ with $k+1$-multiplicity. The resulting expansion is then:
\begin{equation}\lbl{eq:OPElike}
{\Pi}^{\rm HVP}(-Q^2)\underset{{\frac{Q^2}{t_0}\ \ra\  \infty}}{\thicksim}\ -\frac{Q^2}{t_0}\ \sum_{\mathsf{p}=1,2,3,\dots}\ \sum_{k=0,1,2,\dots} \frac{(-1)^{k+1}}{k!}\ \mathsf{\cR}_{\mathsf{p},k}\ \left(\frac{Q^2}{t_0}\right)^{-\mathsf{p}}\ \log^{k}\frac{Q^2}{t_0}\,,
\end{equation}
with the residues $\mathsf{\cR}_{\mathsf{p},k}$ becoming   the coefficients of the large-$Q^2$  asymptotic expansion. These residues  encode complementary information about the hadronic spectral function  to the one provided by the moments in Eq.~\rf{eq:moments}, though they are not so easily accessible to   experimental determination. The terms with one $\log\frac{Q^2}{t_0}$ power in particular, are  generated by the double poles in the r.h.s. of  Eq.~\rf{eq:sse} which arise from the combination  of the poles of $\Gamma(1-s)$ with the simple poles of $\cM^{\rm HVP}(s)$. As we shall see, it is because of the presence of  non-analytic terms in this asymptotic expansion (terms with  $k\ge 1$),   that  a specific transfer theorem is of relevance to HVP.

\end{itemize}
 	
\subsection{Conformal Mapping}

The  framework that we shall be using is the one which   follows from performing a conformal mapping of the full complex $q^2$-plane onto  the unit disc $\vert \omega\vert\le 1$ via the transformation
\be
i\frac{1+\omega}{1-\omega}=\sqrt{\frac{q^2}{t_0}-1}\,.
\ee

\begin{figure}[h!]
\begin{center}
  \includegraphics[scale=1]{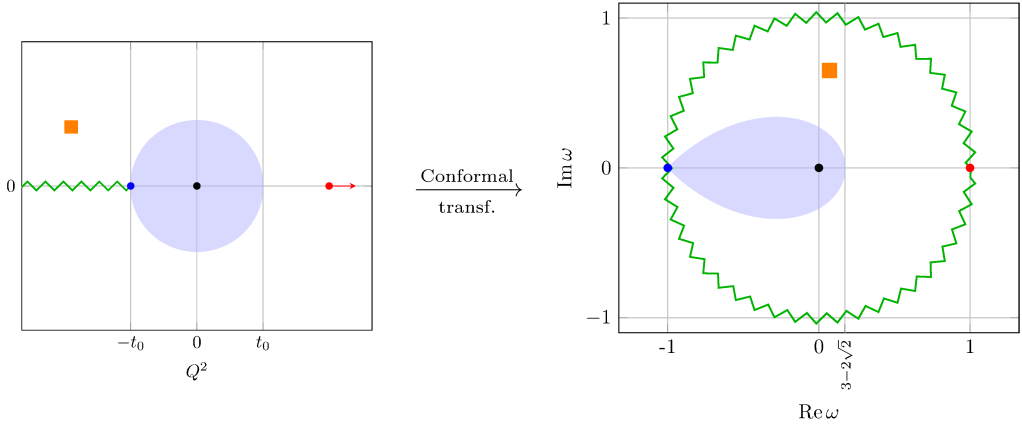}
\caption{Conformal mapping of the $Q^2$-plane in the left into the $\omega$-plane at the right. The green zigzag lines correspond to the branching cut $\frac{Q^2}{t0}<-1$. The black dot represents the $Q^2=0$ point on the left, mapped to $\omega=0$ on the right. The blue dot corresponds to the physical threshold: $Q^2=-t_0$ on the left mapped to $\omega=-1$ on the right. The red dot is the limit $Q^2\rightarrow\infty$ on the left  mapped to $\omega=1$ on the right. The zones in blue are the regions where $\vert Q^2\vert \le t_0$ in the Taylor expansion of $\Pi(Q^2)$: the disc on the left  is mapped to the blue domain in the $\omega$-variable. The orange square shows how an arbitrary point in the complex $Q^2$-plane is transformed in the $\omega$-plane.}  
\lbl{fig:Planes}
\end{center}
\end{figure}

\noi
This mapping, illustrated in Figure~1, relates the Euclidean $Q^2$-variable to the conformal $\omega$-variable as follows:
\be\lbl{eq:confeq}
z\equiv \frac{Q^2}{t_0}
=\frac{4\omega}{(1-\omega)^2} \quad\annd\quad \omega=\frac{\sqrt{1+z}-1}{\sqrt{1+z}+1}\,.
\ee
Under this change of variables, the Taylor series in the r.h.s. of Eq.~\rf{eq:PiExpQ}, becomes a  power series in terms of the dimensionless  $\omega$-variable:
\begin{equation}
  \lbl{eq:PiOmegaExp}
 \Pi^\mathrm{HVP}(-Q^2) \mapsto \Pi^\mathrm{HVP}\left(-\frac{4\omega}{(1-\omega)^2}\right) \underset{|\omega|<1}{\sim} \sum_{n=1}^\infty \Omega_n  \; \omega^n = \sum_{n=1}^\infty \Omega_n  \left(\frac{\sqrt{1+\frac{Q^2}{t_0}}-1}{\sqrt{1+\frac{Q^2}{t_0}}+1} \right)^n\;,
\end{equation}
with coefficients $\Omega_n$~\footnote{ One advantage of expressing $\Pi^\mathrm{HVP}(-Q^2)$ as a   power series in $\omega$ is that it increases considerably the rate of convergence. Both series in $\frac{Q^2}{t_0}$-powers  and in $\omega$-powers converge to the same function; however,  with the same number of terms in their  expansions,  the precision is better if one uses the series in $\omega$ rather than the series in $\frac{Q^2}{t_0}$.} that are  linear combinations of the Mellin moments in Eq.~\rf{eq:moments}:
\be\lbl{eq:OmegaasM}
\Omega_n=\sum_{p=1}^{n}(-1)^p\  4^p\ \frac{\Gamma(n+p)}{\Gamma(2p)\ \Gamma(n+1-p)}\ \cM^{\rm HVP}(1-p)\,,
\ee
and reciprocally
\be
\cM^{\rm HVP}(-n)=\frac{\Gamma(1+n)\ \Gamma\left(\frac{3}{2}+n \right)}{\sqrt{\pi}}\ \sum_{k=1}^{n+1} \frac{(-1)^k\  k}{\Gamma(2+n-k)\ \Gamma(2+n+k)}\ \Omega_k\,. 
\ee
 
\noi
{On the other hand}, the conformal mapping of the asymptotic expansion  in Eq.~\rf{eq:OPElike} when $Q^2 \ra\infty$, results then  in an asymptotic expansion for $\omega\ra 1$:
\be\lbl{eq:OPEomegaR}
{\Pi}^{\rm HVP}\left(-\frac{4\omega}{(1-\omega)^2}\right)\underset{{\omega \ra\ 1}}{\thicksim}\ \sum_{\mathsf{p}=1,2,3,\dots}\ \sum_{k=0,1,2,\dots} \frac{(-1)^{k}}{k!}\ \mathsf{\cR}_{\mathsf{p},k}\ \left(\frac{(1-\omega)^2}{4\omega}\right)^{\mathsf{p}-1}  \log^{k}\frac{4\omega}{(1-\omega)^2}\,,
\ee
and the  transfer theorem becomes relevant because of the  non-analytic $k\neq 0$ terms of this series.

\subsection{The Imaginary Part}

\noi
The terms of the Taylor series in Eq.~\rf{eq:PiOmegaExp} have imaginary parts which can be obtained by analytic continuation taking the following steps:

\begin{enumerate}

\item
First observe that the second degree equation  in $\omega$ in  Eq.~\rf{eq:confeq} has a discriminant
\be
\Delta \equiv 16\left(1+z \right)\,,\quad{\rm where}\quad z\equiv\frac{Q^2}{t_0}\,.
\ee

\item 
For $z>-1$ one has $\Delta>0$ and we get two solutions for $\omega$:
\be
\omega_{+}=\frac{\sqrt{1+z}-1}{\sqrt{1+z}+1}\quad\annd\quad
\omega_{-}=\frac{\sqrt{1+z}+1}{\sqrt{1+z}-1}\,.
\ee
The solution $\omega\equiv\omega_{+}$   is the one in Eq.~\rf{eq:confeq}, which we chose  because it keeps the equivalence of the limits $Q^2\ra 0$ and $\omega\ra 0$. 

\item
For $z<-1$ one has $\Delta<0$, and in this case it is convenient to introduce the energy-momentum squared  variable $\tau$:
\be
\tau\equiv -z=\left(-\frac{Q^2}{t_0}\right)\quad{\rm with}\quad \tau\equiv\frac{t}{t_0}\ge 1\,.
\ee
The  analytic continuation of the $\omega\equiv\omega_{+}$  solution is then 
\be
\frac{i\sqrt{\tau -1}+1}{i\sqrt{\tau-1}-1}\,.
\ee

\item
We therefore have a twofold representation for the conformal $\omega$-variable, either in terms of  $Q^2$ or in terms of  $\tau$:

{\setl 
\bea\lbl{eq:omegas}
  \omega = \begin{cases}
  \frac{\sqrt{1+\frac{Q^2}{t_0}}-1}{\sqrt{1+\frac{Q^2}{t_0}}+1} & \text{when  } \frac{Q^2}{t_0}>-1\,,  \\
  \\
\frac{i\sqrt{\tau-1}+1}{i\sqrt{\tau-1}-1}\equiv  e^{-i\varphi} & \text{when  } \frac{Q^2}{t_0}=-\tau\le -1\,, 
\end{cases}
\eea}

\noi
with the phase $\varphi$  related to the $\tau$-variable as follows~\footnote{This determination takes into account the fact that the real part of $\omega$ (i.e. $\cos \varphi$) can be positive or negative.}:
\begin{equation}
  \varphi = 2 \arctan \left(\frac{\Imm\, \omega}{\Ree\, \omega + |\omega|}\right) = 2 \arctan \left(\frac{1}{\sqrt{\tau-1}}\right)=\pi - 2 \arctan \left(\sqrt{\tau-1} \right)\,.
\end{equation}
Then
\be
\sin\varphi=\frac{2\sqrt{\tau-1}}{\tau} \quad\annd\quad \cos\varphi=\frac{\tau-2}{\tau}\,. 
\ee

\item
For $\vert\omega\vert=1$, the series in Eq.\rf{eq:PiOmegaExp}
 generates  an imaginary part  series: 
\begin{equation}
 \sum_{n=1}^\infty \Omega_n\  \Imm\left(\omega^n \right) = - \sum_{n=1}^\infty \Omega_n \sin (n\varphi) 
\end{equation}
which, in terms of the physical $\tau$-variable, becomes a spectral function series:

{\setl
\bea
\Imm \Pi^\mathrm{HVP}\left(\tau\equiv \frac{t}{t_0}\right) &  = &  -\sum_{n=1}^\infty \Omega_n 
\sin(n\varphi) =-\sum_{n=1}^\infty\Omega_n  \sin\varphi\ {\rm U}_{n-1}(\cos\varphi)  \lbl{eq:fps1} \\
& = & 
-\sum_{n=1}^\infty\Omega_n  \frac{2\sqrt{\tau-1}}{\tau}\ {\rm U}_{n-1}\left(\frac{\tau-2}{\tau}\right)\,, \lbl{eq:fps2}
\eea}

\noi 
where ${\rm U}_{n-1}(\cos\varphi)$ are Chebyshev polynomials of the second kind. 

\end{enumerate}

\noi
In QCD,  the shape of the spectral function at threshold is fixed by lowest order chiral perturbation theory ($\chi$PT):
\begin{equation}\lbl{eq:lochipt}
\frac{1}{\pi} \Imm \Pi^\mathrm{HVP}\left(\tau\equiv \frac{t}{t_0}\right) \underset{\tau\rightarrow1}{=}  \frac{\alpha}{\pi} \frac{1}{12}\left(\tau-1\right)^{\frac{3}{2}} + \mathcal{O}\left[\left(\tau-1\right)^{\frac{5}{2}}\right]\,;
\end{equation}
however,  the behaviour at $\tau\ra 1$ which follows from the expansion:  
\begin{equation}
\sin(n\varphi) \underset{\tau\rightarrow1}{=} -2n(-1)^n \sqrt{\tau-1} + \frac{2}{3}n(1+2n^2)(-1)^n \left(\tau-1\right)^{\frac{3}{2}} + \mathcal{O}\left[\left(\tau-1\right)^{\frac{5}{2}}\right]
\end{equation}
in Eq.~\rf{eq:fps1},
results in a threshold behaviour:
\begin{multline}
\frac{1}{\pi} \Imm \Pi^\mathrm{HVP}\left(\tau\equiv \frac{t}{t_0}\right) \underset{\tau\rightarrow1}{=} \left[\frac{2}{\pi} \sum_{n=1}^\infty \Omega_n n (-1)^n\right] \sqrt{\tau-1} \\
+  \left[-\frac{2}{3\pi} \sum_{n=1}^\infty \Omega_n n(1+2n^2)(-1)^n\right] \left(\tau-1\right)^{\frac{3}{2}} + \mathcal{O}\left[\left(\tau-1\right)^{\frac{5}{2}}\right]\,.
\end{multline}
For consistency with $\chi$PT, this requires two constraints on the $\Omega_n$-coefficients:
\begin{equation}\lbl{eq:thrcons}
 \sum_{n=1}^\infty \Omega_n n (-1)^n =0\quad \text{and} \quad   -\frac{4}{3\pi} \sum_{n=1}^\infty \Omega_n n^3(-1)^n = \frac{\alpha}{\pi} \frac{1}{12}\vert F(t_0)\vert^2\,,
\end{equation}
where $F(t_0)$ denotes the value of the electromagnetic pion form factor at threshold, which encodes the full  hadronic  correction to the lowest order  $\chi$PT result  where $F(t_0) =1$. We shall later explain how to include these two constraints in practice. 

On the other hand,  the spectral function in QCD when $\tau\ra\infty$ is fixed by the asymptotic freedom limit:
\begin{equation}
  \lbl{eq:pQCD}
\frac{1}{\pi}\Imm \Pi^\mathrm{HVP}\left(\tau\equiv \frac{t}{t_0}\right) \underset{\tau \rightarrow \infty}{\sim} \frac{\alpha}{\pi} \frac{N_c}{3} \sum_{\rm quarks}e_{q}^2\,.
\end{equation}
The series in Eq.~\rf{eq:fps2}, however,  does not reproduce this property because for a fixed $n$-term 
\be
{\rm U}_{n-1}\left(\frac{\tau-2}{\tau}\right)
\underset{\tau\ra\infty}{\sim} n-\frac{2}{3}n(n^2 -1)\frac{1}{\tau}+\cO \left(\frac{1}{\tau^2} \right)\,.
\ee
This is an indication that the formal series in Eq.~\rf{eq:fps2} must be a divergent series,  which  is precisely the point that  brings us 
to the relevance of the  transfer theorem  discussed  
 in the next section.

\section{Reconstruction Approximants  of the HVP  Function}

The transfer theorems of Flajolet and Odlyzko~\cite{FOth,FS09} relate the non-analyticity of a given function defined in the unit $\omega$-disc, to the large order coefficients of its Taylor expansion. As already mentioned,   in the  case of the  HVP self-energy,  the non-analyticity  originates in the $\log^{k}(1-\omega)$-terms of the asymptotic series in Eq~\rf{eq:OPEomegaR}. For the  $k=1$ terms in particular:

\begin{multline}
\sum_{\mathsf{p}=1,2,3,\dots}\ \left(-2\ \mathsf{\cR}_{\mathsf{p},1}\right)4^{1-\mathsf{p}}\ \left(\frac{(1-\omega)^2}{4\omega}\right)^{\mathsf{p}-1} \log\left(\frac{1}{1-\omega}\right)\\
\underset{\omega\ra 1}{=}
  \sum_{m=0,1,2,\ldots} \widetilde{\mathcal{R}}_{m,1}\, (1-\omega)^m \log\left(\frac{1}{1-\omega}\right)\,,
\end{multline}
with  coefficients  $\widetilde{\mathcal{R}}_{m,1}$  that are linear combinations of the $\mathsf{\cR}_{\mathsf{p},1}$ residues
\be\lbl{eq: Rcoeffs}
  \widetilde{\mathcal{R}}_{0,1} = -2 \mathcal{R}_{1,1}\;\; \text{and  }\;  \widetilde{\mathcal{R}}_{m,1} = -2 \sum_{\mathsf{p}=2}^{\lfloor\frac{m+2}{2}\rfloor}  \binom{m-\mathsf{p}}{\mathsf{p}-2} 4^{1-\mathsf{p}} \mathcal{R}_{\mathsf{p},1}\quad\foor\quad m\ge 1\,,
\ee 
{and in the pQCD asymptotic freedom limit}
\be
{\widetilde{\mathcal{R}}_{0,1} = -2
\mathsf{\cR}_{\mathsf{1},1}=-2\ \frac{\alpha}{\pi} \frac{N_c}{3} \left(\sum_{\rm quarks}e_{q}^2\right)\,.}
\ee

The appropriate transfer theorem in this case~\footnote{See  Appendix B for a proof, where the general case for $k\ge 1$ is also discussed.} states   that the behaviour of the  $\Omega_n$ coefficients of the Taylor series
\begin{equation}
  \lbl{eq:seriesunitdisk}
 \Pi^\mathrm{HVP}\left(-\frac{4\omega}{(1-\omega)^2}\right) \underset{|\omega|<1}{\sim} \sum_{n>0} \Omega_n \omega^n\,,
\end{equation} 
\underline{at their large-$n$ values},
must be of the form
\be\lbl{eq:largen}
  \Omega_n \underset{n\rightarrow \infty}{\sim}\Omega^{\rm AS}_n = \sum_{j=0}^\infty \sum_{m=0}^\infty  \widetilde{\mathcal{R}}_{m,1} \begin{Bmatrix} m+j \\ m \end{Bmatrix} \frac{(-1)^m\Gamma(m+1)}{n^{m+j+1}} \,,
\ee
where $\begin{Bmatrix} a \\ b \end{Bmatrix}$ are  Stirling numbers of the second kind~\cite{ADAMCHIK1997119}.

The {\it {\it reconstruction approximants}} of the HVP self-energy function consists then in replacing the infinite sums in the   identity:
\be\lbl{eq:fexact}
 \Pi^\mathrm{HVP}\left(-\frac{4\omega}{(1-\omega)^2}\right)  =     \sum_{n=1}^{\infty} (\Omega_n -\Omega_n^\mathrm{AS})\ \omega^n+\sum_{n=1}^\infty \Omega_n^\mathrm{AS}\ \omega^n \,,
\ee
by successive 
{\rm N,L}-functions:
\be\lbl{eq:fn}
 \Pi^\mathrm{HVP}_{\rm N,L}\left(-\frac{4\omega}{(1-\omega)^2}\right)  \doteq     \sum_{n=1}^{\rm N} (\Omega_n -\Omega_{n,{\rm L}}^\mathrm{AS})\ \omega^n\ +\ \sum_{n=1}^\infty \Omega_{n,{\rm L}}^\mathrm{AS}(\omega)\,,
\ee
with $\sum_{n=1}^\infty \Omega_{n,{\rm L}}^\mathrm{AS}(\omega)$ evaluated in terms of a finite sum of polylog $\operatorname{Li}_{l}(\omega)$   functions with $1\le l\le {\rm L}$. These  polylogs result from applying their  definition:   
\be\lbl{eq:polyL}
\operatorname{Li}_{l}(\omega)=\sum_{n=1}^{\infty}\ \frac{\omega^n}{n^l}\,,\quad \operatorname{Li}_{l}(\omega)=\int_0^\omega \frac{d\omega}{\omega}\operatorname{Li}_{l-1}(\omega)\,, \quad \operatorname{Li}_{1}(\omega)=-\log(1-\omega)\,,\quad \vert\omega\vert\le 1\,,
\ee
to the   power  factors $\frac{1}{n^{m+j+1}}$  in the asymptotic  $\Omega_n^\mathrm{AS}$ series in Eq.~\rf{eq:largen}. Setting $m+j+1\doteq l$, we have
\be
\sum_{n=1}^\infty \Omega_{n,{\rm L}}^\mathrm{AS}(\omega)=\sum_{l=1}^{\rm L}\cB_{l}\ \sum_{n=1}^\infty \frac{\omega^n}{ n^l}=\sum_{l=1}^{\rm L}\cB_{l}\ \operatorname{Li}_{l}(\omega)
\ee
where, as shown in the Appendix B, the $\cB_{l}$-coefficients  are linear combinations of the $\widetilde{\mathcal{R}}_{m,1}$-coefficients in Eq.~\rf{eq: Rcoeffs} and hence, linear combinations  of the residues $\cR_{\mathsf{p},1}$:
\begin{equation}
\mathcal{B}_1 = -2 \mathcal{R}_{1,1} \; \text{and   }\; \mathcal{B}_l = \sum_{m=1}^{l-1}  \widetilde{\mathcal{R}}_{m,1} \begin{Bmatrix} l-1 \\ m \end{Bmatrix} (-1)^m\Gamma(m+1)\quad  {\rm with}\quad \cB_{l~{\rm even}}=0 \,.
\end{equation}

\vspace*{0.5cm}

The successive {\it {\it reconstruction approximants}} in Eq.~\rf{eq:fn} are thus defined by  partial N and L sums:
\be\lbl{eq:fNL}
 \Pi^\mathrm{HVP}_{\rm N,L}\left(-\frac{4\omega}{(1-\omega)^2}\right)  \doteq    \sum_{n=1}^{\rm N} \underbrace{(\Omega_n -\Omega_{n,{\rm L}}^\mathrm{AS})}_{\cA(n,{\rm L})}\ \omega^n +  \sum_{l=1}^{\rm L} \cB_{l} \operatorname{Li}_{l}(\omega)\,,
\ee
with  the $\cA(n,{\rm L})$ and $\cB_l$ parameters of the approximation to be fixed either from the Mellin transform of the spectral function {of} the underlying theory, as illustrated with the QED example in Appendix I, or  as in the case of the  MUonE proposal, from a fit  to the data values of the experimental determination of the HVP self-energy function. 
  As the number of  
 N  terms  increases  and the number L of polylog functions  increases, the $ \Pi^\mathrm{HVP}_{\rm N,L}\left(-\frac{4\omega}{(1-\omega)^2}\right)$ approximants 
reconstruct better and better the wanted $\Pi^\mathrm{HVP}\left(-\frac{4\omega}{(1-\omega)^2}\right)$ function. 

We shall later  discuss how to attribute a systematic error to a fixed  $\Pi^\mathrm{HVP}_{\rm N,L}\left(\cdots\right)$ approximant, hence, to its contribution to $a_{\mu}^{\rm HVP}$.

\vspace*{0.5cm}

As compared to other procedures discussed in the literature, the method of {\it reconstruction approximants},  only uses   information provided by the experimental determination of the HVP self-energy in a specific set of values of its argument. It   does not require extra input from other  sources, like LQCD and/or phenomenological models. This offers, therefore, an opportunity for a future MUonE experiment to provide a completely independent  determination of $a_{\mu}^{\rm HVP}$.

\vspace*{1cm}

\subsection{Approximants from Asymptotic Freedom}

The simplest examples of  {\it reconstruction approximants} of the HVP self-energy function  are  the ones with L=1 and a few N terms. 

In QCD,  the leading singular behaviour of the function  $\Pi^{\rm HVP}(-Q^2)$ at large-$Q^2$  is  governed by the first pole at $s=1$ in the singular series  of Eq.~\rf{eq:largen}, with a residue $\mathsf{\cR}_{\mathsf{1},1}$ known from  the asymptotic freedom limit behaviour:
\be
  \lbl{eq:PerturbativeQCD}
\Pi^\mathrm{HVP}(-Q^2) \underset{Q^2 \rightarrow \infty}{\sim} -\mathsf{\cR}_{\mathsf{1},1}\  \log  \left(\frac{Q^2}{t_0}\right)\,,\quad {\rm and}\quad \mathsf{\cR}_{\mathsf{1},1}=\frac{\alpha}{\pi} \frac{N_c}{3} \left(\sum_{\rm quarks}e_{q}^2\right)\,.
\ee
There are corrections to this limit generated by the $\alpha_{\rm QCD}(Q^2)$ power series of pQCD. We shall not include them here although, if necessary, they could also be taken into account using the results discussed in  Appendix B.

In the conformal $\omega$-disc, asymptotic freedom produces the non-analytic term:
\begin{equation}\lbl{eq:firstlog}
\Pi^\mathrm{HVP}\left(-\frac{4\omega}{(1-\omega)^2}\right) \underset{\omega\rightarrow 1}{\sim}   -2\ \mathsf{\cR}_{\mathsf{1},1}\ \log \left(\frac{1}{1-\omega}\right)
\end{equation}
and therefore, according to Eqs.~\rf{eq:largen} and \rf{eq:fNL},
\begin{equation}\lbl{eq:omegaas}
  \Omega_n \underset{n\rightarrow \infty}{\sim} \Omega_n^{\mathrm{AS}} =   -2\ \mathsf{\cR}_{\mathsf{1},1}\  \frac{1}{n}\quad\annd\quad \cA(n,1)\equiv\Omega_{n} +2\ \cR_{1,1}\ \frac{1}{n}\,. 
\end{equation}
In this case,  the {\it reconstruction approximants} of Eq.~\rf{eq:fNL}  are rather simple. They consist in evaluating the successive N-sums:
\be  \lbl{eq:PiApprox}
\Pi^\mathrm{HVP}_{{\rm N},1}(-Q^2)= \sum_{n=1}^{\rm N} \cA(n,1)  \; \left(\frac{\sqrt{1+\frac{Q^2}{t_0}}-1}{\sqrt{1+\frac{Q^2}{t_0}}+1}  \right)^n 
+  2\ \mathsf{\cR}_{\mathsf{1},1}\ \log \left(1-\frac{\sqrt{1+\frac{Q^2}{t_0}}-1}{\sqrt{1+\frac{Q^2}{t_0}}+1}   \right)\,,
\ee
 with only one reconstruction term which in this case is  the leading polylog $\operatorname{Li}_{1}(\omega)=-\log(1-\omega)$ function.

The application of this result to the MUonE-experiment consists then in using the sum of functions above as a  ``first set of  approximants'' to the full integrand in the r.h.s. of Eq.~\rf{eq:LdeR}. More precisely, the {\it reconstruction approximants} in this case  are:  

{\setl
\bea\lbl{eq:firststep}
(1-x)\ \Pi^{\rm HVP}_{{\rm N},1}\left( -\frac{x^2}{1-x}m_{\mu}^2\right)  & \doteq &  (1-x)\left\{\sum_{n=1}^{\rm N} \cA(n,{\rm 1})  \ \left(\frac{\sqrt{1+\frac{x^2}{1-x}\frac{m_{\mu}^2}{t_0}}-1}{\sqrt{1+\frac{x^2}{1-x}\frac{m_{\mu}^2}{t_0}}+1}  \right)^n \right.\nn  \\
& + & \left. 2\ \mathsf{\cR}_{\mathsf{1},1}\  \log \left(1-\frac{\sqrt{1+\frac{x^2}{1-x}\frac{m_{\mu}^2}{t_0}}-1}{\sqrt{1+\frac{x^2}{1-x}\frac{m_{\mu}^2}{t_0}}+1}   \right)\right\}\,,
\eea}

\noi
with unknown coefficients $\cA(n,{\rm 1})$, which are free parameters to   be  fixed  from a linear fit to the experimental data in the $x$-window where the quality of the data is best. Using the values of the $\cA(n,{\rm 1})$ parameters thus obtained, Eq.~\rf{eq:LdeR} provides then the way  to obtain  the corresponding approximate  evaluations of the HVP contribution to the muon anomaly.  

As discussed before, the approximants in Eq.~\rf{eq:PiApprox} have  imaginary parts which give corresponding  ``effective spectral series approximants''. The power terms in Eq.~\rf{eq:firststep}  generate imaginary parts similar to those given in Eq.~\rf{eq:fps2} except that now the series are from $n=1$ to a finite N-value and  the $\cA(n,1)$ coefficients now  replace the $\Omega_n$. 
 The imaginary part of the logarithmic term in  Eq.~\rf{eq:firststep} gives 
\be
\Imm\left[\log(1-\omega)\right]=\Imm\left[\log\left(\frac{2}{\tau}-i\frac{2\sqrt{\tau-1}}{\tau}\right)\right]=-\arctan \left(\sqrt{\tau-1} \right)\,,
\ee
and therefore, the ``effective spectral functions'' associated to the approximants
$\Pi^\mathrm{HVP}_{{\rm N},1}(-Q^2)$ in Eq.~\rf{eq:PiApprox} are then  given by the N-sums (recall that $\tau=\frac{t}{t_0}$):
\be\lbl{eq:specN2}
\frac{1}{\pi}\Imm\Pi^{\rm HVP}_{{\rm N},1}(t)  =\frac{2}{\pi}  \left\{-\ \frac{\sqrt{\tau-1}}{\tau}\  \sum_{n=1}^{\rm N}\cA(n,{\rm 1}) \ {\rm U}_{n-1}\left(\frac{\tau-2}{\tau}\right)  +   \mathsf{\cR}_{\mathsf{1},1}\ \arctan(\sqrt{\tau-1})\right\} \theta(\tau -1)\,.
\ee
To this, one can also apply the 
 constraints in Eq.~\rf{eq:thrcons}  
which can be easily adapted  to the threshold expansion of the r.h.s. of Eq.~\rf{eq:specN2} i.e.~\footnote{It is not necessary to fix the value of $F(t_0)$ from phenomenology. It can also be taken as a free parameter in the fit.}  
\be\lbl{eq:conssimp}
0 =\sum_{n=1}^{\rm N} \cA(n,{\rm 1}) n (-1)^n  +  \mathsf{\cR}_{\mathsf{1},1} \quad \annd\quad 
\frac{\alpha}{\pi} \frac{1}{12}\vert F(t_0)\vert^2 =-\frac{4}{3\pi} \sum_{n=1}^{\rm N} \cA(n,{\rm 1}) n^3(-1)^n\,, \quad{\rm N}\ge 2 \,.
\ee

\subsection{Asymptotic freedom  and Lowest Order $\chi$PT}

The case L=1 with N=2 is particularly interesting because, quite remarkably, the parameters $\cA(1,1)$ and $\cA(2,1)$ can then  be fixed from  asymptotic freedom    and lowest order $\chi$PT alone. In this case, the two constraints in Eq.~\rf{eq:conssimp}, with  $F(t_0)=1$,  result in the values (in $\frac{\alpha}{\pi}$ units)
\be\lbl{eq:A1A2}
\cA(1,{\rm 1}) =2.156\dots\quad \annd \quad \cA(2,{\rm 1}) =0.2450\dots\ \,,
\ee
and the shape of the muon anomaly integrand for these values is the one shown in Fig.~\rf{fig:anaschi}.

\begin{figure}[!ht]
\begin{center}
\hspace*{-1cm}\includegraphics[width=0.60\textwidth]{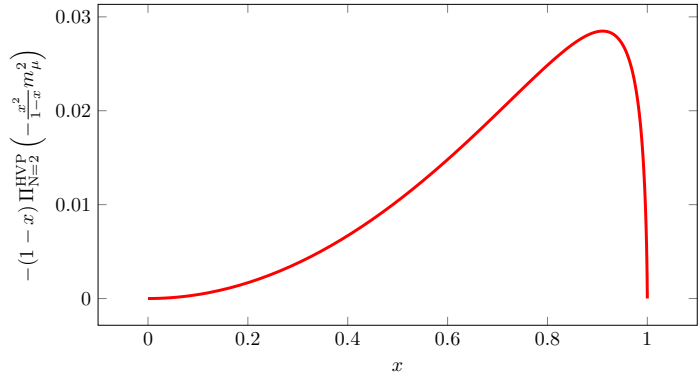} 
\caption{\lbl{fig:anaschi} Shape of the $x$-integrand in Eq.~\rf{eq:LdeR} in $\frac{\alpha}{\pi}$ units, corresponding to the reconstruction approximant: ${\rm N}=2$, ${\rm L}=1$.}
\end{center}
\end{figure}

\noi
The resulting value for the HVP contribution to the muon anomaly  is
\be\lbl{eq:anN2}
a_{\mu}^{\rm HVP}\Big|_{{\rm N}=2}=(6527.12\dots)\times 10^{-11}\,,
\ee
which reproduces the center value of the  dispersive data -driven evaluations in Eq.~\rf{eq:disperevs} at the 6\% level (quite encouraging~\footnote{Not yet the accuracy that one wants of course, but already  better than some of the  LQCD evaluations.}). We wish to emphasize the fact that only two rigorous  limits of QCD have been used to obtain this result: the short-distance asymptotic freedom limit and the long-distance lowest order $\chi$PT limit.

\noi
\begin{figure}[!ht]
\begin{center}
\hspace*{-1cm}\includegraphics[width=0.60\textwidth]{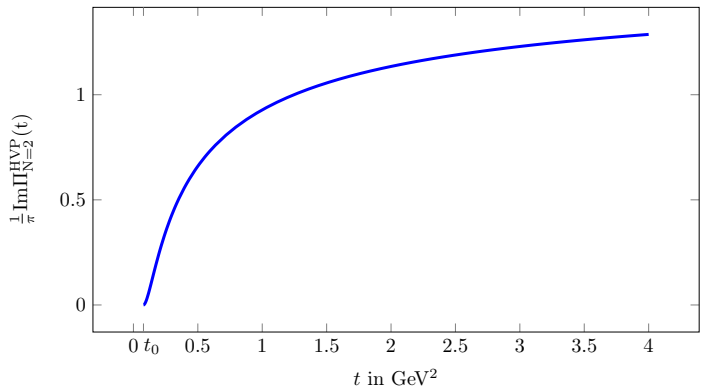} 
\caption{\lbl{fig:specN2} Shape of the spectral function for ${\rm N}=2$ in  $\frac{\alpha}{\pi}$ units.}
\end{center}
\end{figure}

The shape of the "effective spectral function" in Eq.~\rf{eq:specN2}  associated to the $\Pi^\mathrm{HVP}_{{\rm N}= 2}(-Q^2)$ approximant in Fig.~\rf{fig:anaschi} is shown in Fig.~\rf{fig:specN2}. We insist on the fact that, contrary to the Euclidean shape in Fig.~\rf{fig:anaschi} which is expected to be a local approximation at each $x$-value, the corresponding ``effective spectral function'' is not a  locally dual approximation at a fixed $t$-value of the physical spectrum. Only  moments and/or weighted integrals of this ``effective spectral function'' can be considered as good approximants~\footnote{  
The dispersive representation in Eq.~\rf{eq:BMSR} is a weighted integral of the spectral function and the result which follows from inserting the ``N=2 effective spectral function'' in this representation   gives, as expected, the same value as the one in Eq.~\rf{eq:anN2} using the Euclidean representation.}. However, as more and more  
 N-power terms  and L-polylog functions are taken into account  in the reconstruction  approximants of Eq.~\rf{eq:fNL} , the corresponding
"effective spectral functions" are expected to become more and more locally dual to the physical spectral function.  

The reason why we only consider  the functions in Eq.~\rf{eq:PiApprox} as a "first set" of {\it reconstruction approximants},  is because they only include the non-analyticity in the conformal $\omega$-domain associated to asymptotic freedom, i.e. the leading $s=1$ singularity of the hadronic Mellin transform. In  the next section we shall show how to construct  {\it reconstruction approximants}, adapted to the MUonE-experiment, when one also  includes higher order singularities.

\section{Reconstruction Approximants for the MUonE Proposal}

In the QED example discussed in Appendix A,   there is only one mass scale, the fermion mass ${\rm M}$.  The singular pattern of the two-point function is therefore  rather simple.  In the case of the electromagnetic interactions of hadrons, one obviously  expects to have more undergoing  mass scales and, therefore, a more complicated singular pattern. In fact, in QCD, the operator product expansion (OPE)  applied to $\Pi^{\rm HVP}(-Q^2)$  shows the existence of  other mass scales  than quark masses. They appear as vacuum expectation values of local colour singlet operators~\cite{SVZ}, like e.g. the gluon condensate. A possible option could have been to apply the transfer theorem method   using  the OPE contributions as an input. These OPE  contributions, however, are   poorly known both from phenomenology and  theory where  the separation of perturbative and non-perturbative scales is  problematic~\footnote{For a recent discussion, where earlier references can also be found,  see e.g. ref~\cite{Jamin}.}. Concerning the application to the MuonE experiment, we are therefore more inclined to apply the {\it reconstruction  approximants} defined in Eq.~\rf{eq:fNL}, using input from experiment alone.  
 More precisely, in units of $\frac{\alpha}{\pi}$, the approximants to be considered for the integrand in the Feynman-$x$ representation given in  Eq.~\rf{eq:LdeR} are:

\bea\lbl{eq:QCDN}
-(1-x)\ \Pi_{\rm N,L}^{\rm HVP}\left( -\frac{x^2}{1-x}m_{\mu}^2\right) &=  -(1-x)\left\{\sum_{n=1}^{\rm N} \cA(n,{\rm L})  \ \left(\frac{\sqrt{1+\frac{x^2}{1-x}\frac{m_{\mu}^2}{t_0}}-1}{\sqrt{1+\frac{x^2}{1-x}\frac{m_{\mu}^2}{t_0}}+1}  \right)^n \right.\nn  \\
& +\left. \sum_{p=1}^{\left\lfloor \frac{{\rm L}+1}{2}\right\rfloor}\cB(2p-1)\ {\rm Li}_{2p-1}\left(\frac{\sqrt{1+\frac{x^2}{1-x}\frac{m_{\mu}^2}{t_0}}-1}{\sqrt{1+\frac{x^2}{1-x}\frac{m_{\mu}^2}{t_0}}+1}  \right)\right\}\,,
\eea
where the terms in the second line, similarly to the QED example discussed in Appendix A,  correspond to the  Polylog series  of Eq.~\rf{eq:fNL}. However, contrary to the QED example where the coefficients $\cA(n,{\rm L})$ and $\cB(2p-1)$ {are} fixed by the theory, they are here free parameters,  to   be  fixed  from a linear fit of the  functional approximants to the experimental data in the $x$-window where the quality of the data is best.

We finally comment on a technical simplification  that we have made in the derivation of   the {\it reconstruction approximants} above: the fact that we have restricted the singular behaviour of the Mellin transform of the physical hadronic spectral function to poles of multiplicity one at most. In other words,  the non-analytic terms generated by the singular series in Eq.~\rf{eq:OPElike} has been limited to  terms with $k=1$, which corresponds to terms with one power of $\log Q^2$ at most in the large-$Q^2$ expansion. As shown in  Appendix B,  it is possible to generalize the {\it reconstruction approximants} so as to include the  effect  of $k>1$ terms; but this is at the expense of introducing more parameters  and extra derivatives of the ${\rm Li}_{2\mathsf{p}-1}$ functions. Because of the  complexity involved, we have not considered this generalization in this paper. It is reassuring, however, to know that the assumption of simple poles in the Mellin transform   is certainly satisfied, not only in QED, but also in many phenomenological  models  of the hadronic spectral function; in particular for  superpositions of Breit-Wigner-- like terms. In all these cases,  
the transfer theorem guaranties that, as the number of N-terms and polylog L-like terms in Eq.~\rf{eq:QCDN} increases, with more and more data points    used in the fits of  the $\cA(n,{\rm L})$ and $\cB(2p-1)$ parameters, the  approximants converge to the physical Euclidean-integrand in its full range.   

As in the QED example discussed in Appendix A, the approximants in Eq.~\rf{eq:QCDN} have imaginary parts that can be interpreted as ``effective spectral function approximants'' associated to the corresponding Euclidean approximants. They can be evaluated following the same steps as in the QED example, with the result  ($\tau=\frac{t}{t_0}$):

\bea\lbl{eq:spectNQED}
\lefteqn{
\frac{1}{\pi}\Imm\Pi_{\rm N,L}^{\rm HVP}(\tau)  = \frac{\alpha}{\pi}\left\{  -\frac{1}{\pi}\ \frac{ 2\sqrt{\tau-1}}{\tau}\  \sum_{n=1}^{{\rm N}}\cA(n,{\rm L}) \ {\rm U}_{n-1}\left(1-\frac{2}{\tau}\right)\right.} \nn \\ & + & \left.  \sum_{p=1}^{\left\lfloor \frac{L+1}{2}\right\rfloor}\cB(2p-1)\  \operatorname{B}_{2p-1}\left(\frac{\pi -2\arctan(\sqrt{\tau-1})}{2\pi}\right)\right\}\theta(\tau-1)\,,
\eea
where  $\operatorname{B}_{2p-1}(\cdots)$  are Bernoulli polynomials.
As already mentioned, these ``effective spectral function approximants'' are  only globally dual  to the physical spectral function. Only their moments  are to be considered as good approximants to the physical moments.  When introduced in the dispersive integral  representation in Eq.~\rf{eq:BMSR}, which is a weighted integral of the spectral function, they reproduce  the same values for the anomaly as those resulting from the Euclidean integration of the $\Pi_{\rm N,L}^{\rm HVP}\left( -\frac{x^2}{1-x}m_{\mu}^2\right)$ approximants in Eq.~\rf{eq:LdeR}.  

\vspace*{0.5cm}

The threshold constraints discussed earlier in Eqs.~\rf{eq:thrcons} and \rf{eq:conssimp}, when adapted to the threshold expansion of the r.h.s. in Eq.~\rf{eq:spectNQED}, and keeping  terms up to L=5,  are:

\be\lbl{eq:th1}
0 = \sum_{n=1}^\infty \mathcal{A}(n,5)n(-1)^n +\frac{5}{3} - \frac{\pi^2}{12} \mathcal{B}(3) - \frac{7\pi^4}{720} \mathcal{B}(5)\,,
\ee

and

\begin{equation}\lbl{eq:th2}
-\frac{3\pi}{4} \chi_{\frac{3}{2}} = \sum_{n=1}^\infty \mathcal{A}(n,5)(-1)^n n^3 - \frac{1}{2} \mathcal{B}(3) - \frac{\pi^2}{12}\mathcal{B}(5)\,,
\end{equation}
where as in Eq.~\rf{eq:conssimp}  $\chi_{\frac{3}{2}} \equiv \frac{\alpha}{\pi} \frac{1}{12}\vert F(t_0)\vert^2$; but 
$\chi_{\frac{3}{2}}$  can also be taken as a free parameter in the fit.

 The systematic errors of the {\it reconstruction approximants} in Eq.~\rf{eq:QCDN} are given by  the terms   not included in the finite sums which define a specific approximant  i.e., for a fixed L, the systematic error $ \cE_{{\rm N},{\rm L}}(\omega)$ is:
\be
 \cE_{{\rm N},{\rm L}}(\omega)=\sum_{n={{\rm N}+1}}^{\infty} \cA(n,{\rm L})\ \omega^n,
\ee
with the $\cA(n,{\rm L})$ coefficients restricted by the transfer theorem to  behave as
\be\lbl{eq:behA}
\cA(n,{\rm L})\  \underset{n\rightarrow \infty}=\ \cO\left(\frac{1}{n^{{\rm L}+1}} \right)\,,
\ee
which implies that for $n$ sufficiently large, say $n>{\rm N}^*$, and  $\cC_{\rm N^*}$ a constant a priory unknown, 
\be
n^{{\rm L}+1}\vert\cA(n,{\rm L})\vert  \underset{\ n > {\rm N^*}}<  \cC_{\rm N^*}\,.
\ee
This results in an upper bound for the systematic error
\be\lbl{eq:errorbound}
\vert \cE_{{\rm N}^*,{\rm L}}(\omega) \vert\le \left\vert\cC_{\rm N*} \sum_{n={\rm N}^{*}+1}^\infty\frac{\omega^n}{n^{{\rm L}+1}}\right\vert= \left\vert
\cC_{\rm N^*}\  \omega^{{\rm N^*}+1}\ \Phi(\omega,{\rm L}+1,{\rm N^*}+1)\right\vert\equiv \Delta_{\rm error}(\omega, {\rm L},{\rm N^*},\cC_{\rm N^*})\,,
\ee
where $\Phi(\omega,{\rm L}+1,{\rm N^*}+1)$ is  the     Hurwitz-Lerch transcendental function. 
The propagation of this error in the 
 integral which gives the HVP determination of $a_{\mu}^{\rm HVP}$, fixes then an upper bound to  the systematic error of the anomaly evaluation:
\be\lbl{eq:erroran}
\cE\left[a_{\mu}^{\rm HVP}\right]_{{\rm N^*},{\rm L}}  =\frac{\alpha}{\pi}\int_0^1 dx\ (1-x)\ \Delta_{\rm error}(\omega, {\rm L},{\rm N^*},\cC_{\rm N^*})\,,\quad{\rm with}\quad \omega=\frac{\sqrt{1+\frac{x^2}{1-x}\frac{m_{\mu}^2}{t_0}}-1}{\sqrt{1+\frac{x^2}{1-x}\frac{m_{\mu}^2}{t_0}}+1}\,.
\ee
It remains to be seen how to estimate in practice the values of the  ${\rm N^*}$ and $\cC_{\rm N^*}$ parameters, when the only information one has about the underlying dynamics is the one provided by the $x$-data points which have been used to fit the $\cA(n,{\rm L})$ and $\cB(2p-1)$ parameters of the approximant.
In the next subsection, we illustrate the  procedure to follow, with an   example where the ``data'' are obtained  from  a simple phenomenological model of the hadronic spectral function.

\section{Illustration with a Phenomenological Model}

\begin{figure}[!ht]
\begin{center}
\hspace*{-1cm}\includegraphics[width=0.60\textwidth]{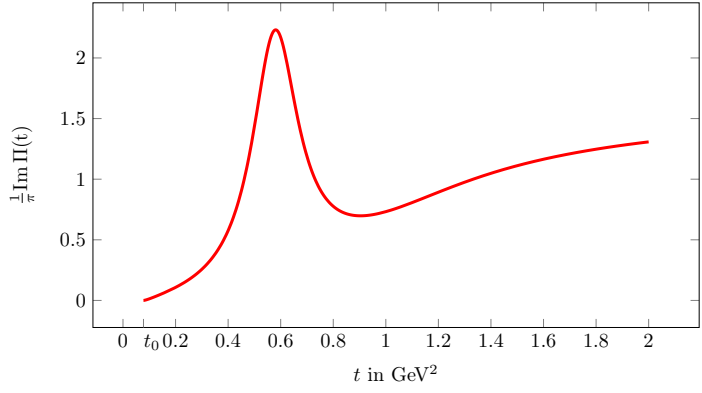} 
\caption{\lbl{fig:modelsp} The model spectral function in Eq.~\rf{eq:SFM} for $t_c =1~\GeV^2$ and $\Delta=0.5~\GeV^2$ in $\frac{\alpha}{\pi}$-units.}
\end{center}
\end{figure}

The spectral function of the model we have adopted  is  inspired from $\chi$PT and  phenomenology~\footnote{It is a simplified version of  phenomenological spectral functions discussed in the literature, see e.g. refs.~\cite{PP,CHK21} and references therein.}:
\noi

\be\lbl{eq:SFM}
\frac{1}{\pi}\Imm\Pi^{\rm HVP}_{\rm model}(t)=\frac{\alpha}{\pi}\left(1-\frac{4 m_{\pi}^2}{t}\right)^{3/2} \left\{ 
\frac{1}{12}\vert F(t)\vert^2 +
\sum_{\rm quarks}e_{q}^2\ \  \Theta(t,t_{c},\Delta)\right\}\theta(t-4m_{\pi}^2)\,.
\ee 
 It has a Breit-Wigner--like modulous squared form factor 
\be\lbl{eq:ff2}
\vert F(t)\vert^2=\frac{M_{\rho}^4}{(M_{\rho}^2-t)^2 +M_{\rho}^2\  \Gamma(t)^2}\,,
\ee
with an  energy dependent width:
\be
\Gamma(t)=\frac{M_{\rho} t}{96\pi f_{\pi}^2}\left[\left(1-\frac{4 m_{\pi}^2}{t}\right)^{3/2}\theta(t-4m_{\pi}^2)+\frac{1}{2}\left(1-\frac{4 M_{k}^2}{t}\right)^{3/2}\theta(t-4M_{k}^2)
\right]\,;
\ee
plus a function
\be
\Theta(t,t_{c},\Delta)=\frac{\frac{2}{\pi}\arctan\left(\frac{t-t_{c}}{\Delta}\right)-\frac{2}{\pi}\arctan\left(\frac{t_{0}-t_{c}}{\Delta}\right)}{1-\frac{2}{\pi}\arctan\left(\frac{t_0-t_{c}}{\Delta}\right)}\,,
\ee
with two arbitrary parameters $t_c$ and $\Delta$. This function has been added so as to smoothly match the low energy phenomenological  spectrum of the model  to  the pQCD asymptotic continuum generated by the sum of quark flavors. 
The shape of this spectral function, using the  physical central values for $m_{\pi}$, $M_k$,  $M_{\rho}$, $f_{\pi}=93.3~\MeV$, and  $t_c =1~\GeV^2$, $\Delta=0.5~\GeV^2$ with   
$\sum_{\rm quarks}e_{q}^2=\frac{5}{3}$,  is shown in Fig~\rf{fig:modelsp}.
\begin{figure}[!ht]
\begin{center}
\hspace*{-1cm}\includegraphics[width=0.60\textwidth]{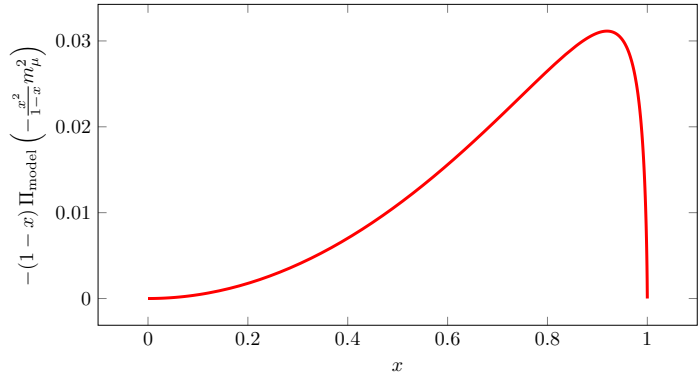} 
\caption{\lbl{fig:modeleuc} Plot of the Euclidean integrand in Eq.~\rf{eq:eucint} for $t_c =1~\GeV^2$ and $\Delta=0.5~\GeV^2$ in $\frac{\alpha}{\pi}$-units.}
\end{center}
\end{figure}

The Euclidean self-energy function of the model with these parameters is given by  the integral
\be
  \Pi_{\rm model}(Q^2) \equiv - \int_{t_0}^\infty \frac{dt}{t} \frac{Q^2}{t+Q^2} \frac{1}{\pi} \Imm\Pi_\mathrm{model}^{\rm HVP}(t)\,,\quad Q^2\equiv \frac{x^2}{1-x}m_{\mu}^2\,.
\ee
It provides a model example to obtain ``data'' at different values of  the Feynman $x$-variable.  
The  contribution to the muon anomaly in this model  is~\footnote{Only central values are used in the model illustration of this section.}
\be\lbl{eq:eucint}
a_{\mu}^{\rm HVP}({\rm model})=-\frac{\alpha}{\pi}\int_0^1 dx\ (1-x)\ \Pi_{\rm model}\left(\frac{x^2}{1-x}m_{\mu}^2\right)\ =\ 6992.4\times 10^{-11}\,,
\ee
and the shape of the integrand is shown in Fig.~\rf{fig:modeleuc}.

We next propose  to fix the values of the  parameters $\cA(n,{\rm L})$ and $\cB(2p-1)$ of   the {\it reconstruction approximants} in Eq.~\rf{eq:QCDN},  
from a linear fit  to  ``data-points'' provided by the $\Pi_{\rm model}(Q^2)$  function. We choose for that  a set of $x$-points in the interval $0.23\le x\le 0.93$ in fifty  equal steps; a choice motivated by the $x$-region   where the MUonE experiment expects to have the best quality of
measurements~\cite{CC}. We have used the Mathematica {\bf nlm} code to do the linear fits, with the following results: 

\subsection{ Approximants with  L=1 only }

These are the ``first step approximants'' that were introduced in  Eq.~\rf{eq:firststep}.  
	The   results  for $a_{\mu}^{\rm HVP}({\rm N})$, for a number  of N terms with the function L=1, are the black dot points in  Fig.~\rf{fig:L1errors} with an estimate, discussed below,  of the systematic error bars  included. The vertical scale in the figure corresponds to a choice of 0.1\% accuracy, attributed to the model value of the anomaly represented  by the horizontal red dashed line. 
\begin{figure}[!ht]
\begin{center}
\hspace*{-0.5cm}\includegraphics[width=0.70\textwidth]{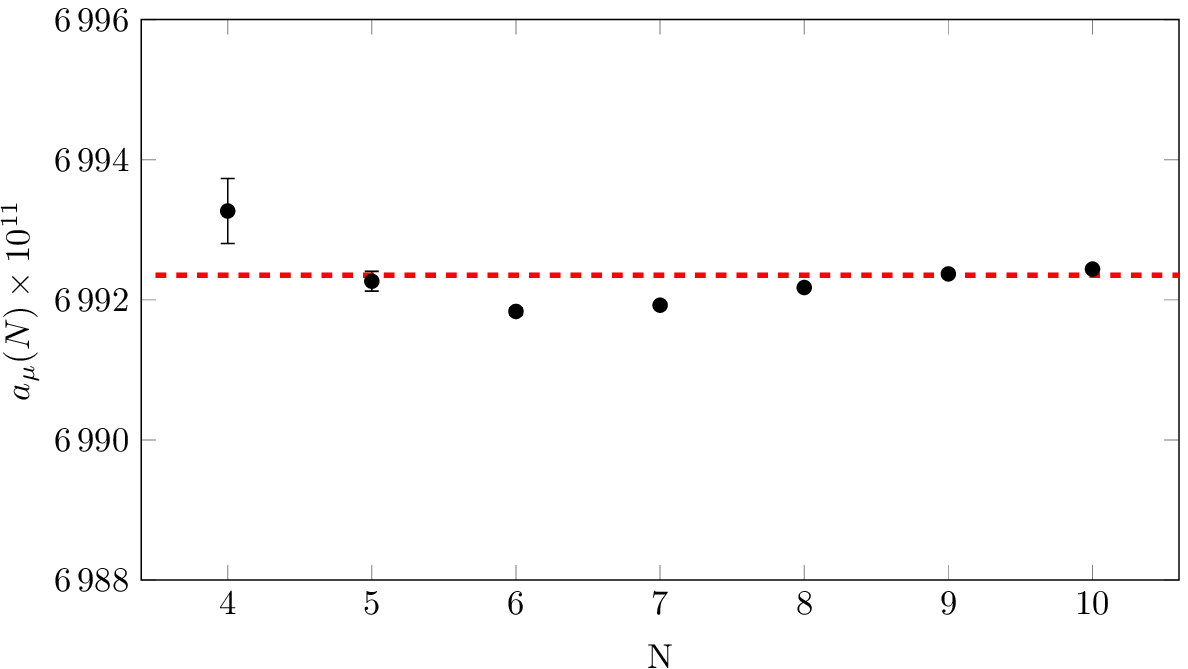} 
\caption{\lbl{fig:L1errors}  $a_{\mu}^{\rm HVP}({\rm N})$ results with estimated  systematic error bars. The horizontal red dashed line is the $a_{\mu}^{\rm HVP}$ value of the model. }
\end{center}
\end{figure}
	
\hspace*{1cm} The estimate of the systematic error bars is based	
on the discussion  at the end of Section IV (Eqs.~\rf{eq:errorbound} and \rf{eq:erroran} in particular): 
 
{\bf i)} For a fixed N, we choose ${\rm N^*} ={\rm N}+1$ and  the constant $\cC_{\rm N^*}$ evaluated by the quadratic mean of the previously evaluated $\cA(n,{\rm L}=1)$ coefficients:
	
	\be\lbl{eq:cstar}
\cC_{\rm N^*}\equiv \sum_{n=1}^{\rm N}\frac{1}{{\rm N}}\sqrt{\left\vert\cA(n,{\rm L}=1)\right\vert^2}\,.
	\ee

	{\bf ii)} The systematic error attributed to the evaluation of the anomaly $a_{\mu}^{\rm HVP}({\rm N})$  is then given by the integral in Eq.~\rf{eq:erroran} with ${\rm N^*}$ and $\cC_{\rm N^*}$ fixed as previously explained.
	
	As shown in Fig.~\rf{fig:L1errors} (notice the vertical scale in the figure), for N=4,5,..,10 the  results are very stable and  reproduce the model value in Eq.~\rf{eq:eucint}, represented by the horizontal dashed red line, to an excellent accuracy.
 Beyond ${\rm N}\sim 10$,  the results become unstable and larger, which is an indication that there is a limit on the number of N-terms one can take in the {\it reconstruction approximants}  when keeping only the leading L=1  function.

\subsection{Reconstruction Approximants  with Threshold Constraints}

The {\it reconstruction approximants} in this case are those in Eq.~\rf{eq:QCDN} where,  for each fixed number of N-power terms,  we add to the leading  ${\rm L}=1$ function  the contribution from the next  two polylog functions with coefficients $\cB(3)$ and $\cB(5)$. The parameters to be fixed  are then
\be
 \cA(n,{\rm L})\quad {\rm with}~n=1,2,3,\dots,{\rm L}\le 5\qquad\annd\qquad \cB(3)\,,\ \cB(5)\,;
\ee
restricted by the threshold constraints  in Eqs.~\rf{eq:th1} and \rf{eq:th2}. The two linear constraint equations can then be used to fix the $\cB(3)$ and $\cB(5)$ parameters in terms of the $\cA(n,{\rm L})$ and $\chi_{\frac{3}{2}} \equiv \frac{\alpha}{\pi} \frac{1}{12}\vert F(t_0)\vert^2$,  which are then  left as  free parameters for a fit to the data. 

\begin{figure}[!ht]
\begin{center}
\hspace*{-0.5cm}\includegraphics[width=0.70\textwidth]{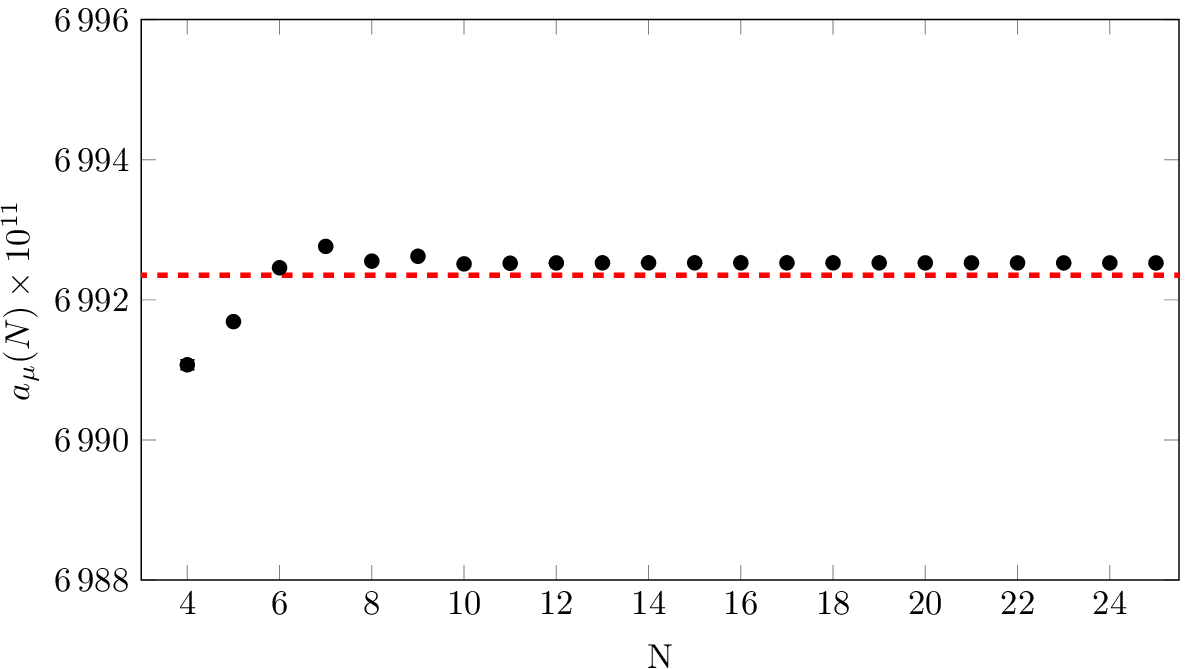} 
\caption{\lbl{fig:L5errors} $a_{\mu}^{\rm HVP}({\rm N},{\rm L}=5)$ results with estimated systematic error bars. The horizontal red dashed line is the $a_{\mu}^{\rm HVP}$ value of the model. }
\end{center}
\end{figure}

\noi
	The   results  for $a_{\mu}^{\rm HVP}({\rm N},{\rm L}=5$), for a number  of N terms with three L-functions  are the black dot points in  Fig.~\rf{fig:L5errors}. The estimated systematic error bars are given by $\Delta_{\rm error}(\omega, {\rm L},{\rm N^*},\cC_{\rm N^*})$ in Eqs.~\rf{eq:errorbound} and \rf{eq:erroran}, with the choice of $\cC_{\rm N^*}$ the same as in Eq.~\rf{eq:cstar} but for ${\rm L}=5$. The vertical scale is the same as in Fig.~\rf{fig:L1errors}, which corresponds to a choice of 0.1\% accuracy, attributed to the model value of the anomaly represented  by the horizontal red dashed line. As shown in 
Fig.~\rf{fig:L5errors} the results remain, very accurate, and stable  up to a very high number of terms.

\hspace*{1cm} As already discussed before, the {\it reconstruction approximants}  have imaginary parts; in this case  given by Eq.~\rf{eq:spectNQED}. They can be considered as ``globally equivalent'' spectral function approximants.   Figure~\rf{fig:specteq} shows, in blue,  the shape  of the ``equivalent'' spectral function approximant corresponding to the choice N=20. Although it is not ``locally dual'' to the model spectral function in red, its weighted integral using Eq.~\rf{eq:BMSR}, reproduces the same value for $a_{\mu}^{\rm HVP}(20,5)$ as using the $x$-Feynman  representation in Eq.~\rf{eq:LdeR}.

\vspace*{1cm}

\begin{figure}[!ht]
\begin{center}
\hspace*{-0.5cm}\includegraphics[width=0.80\textwidth]{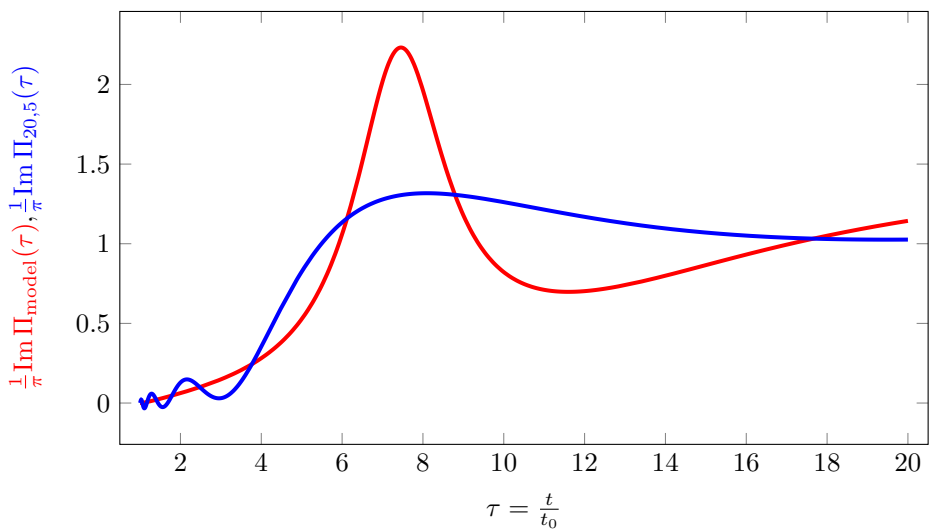} 
\caption{\lbl{fig:specteq} "Equivalent" spectral function approximant in blue,  for N=20 and L=5. In red the model spectral function.}
\end{center}
\end{figure}


\section{Conclusions and Outlook}

We conclude that {\it Reconstruction Approximants}, based on the work of Flajolet and Odlyzko~\cite{FOth}, provide an excellent way to extrapolate the determination of two-point functions in regions  neither  covered by model independent evaluations nor by precise enough experimental determinations. In the Appendix I, we have illustrated the underlying theory of reconstruction with the example of vacuum polarization in QED at the one loop level,  where all the steps have been made analytically. Our main purpose, however,  has been to show how to apply these approximants  to extend the evaluations of the HVP self-energy to the  $x$-regions where the MUonE proposal will  not have access  with sufficient accuracy. We have illustrated how to do this in practice with the help of a  phenomenological model. The results, shown in Figs.~\rf{fig:L1errors} and \rf{fig:L5errors}, are extremely encouraging. They don't include, however, the statistical errors of an experiment, neither the optimal choice of the data points to perform a fit. This is something beyond the scope of this paper. 

Based on the work described above,     we believe  that the  use of {\it reconstruction approximants} enhances  the interest  of the MUonE proposal as an independent and competitive way to measure $a_{\mu}^{\rm HVP}$. As compared to other methods suggested in the literature, {\it reconstruction approximants}  don't require extra input from either phenomenological estimates or LQCD evaluations. This gives the possibility that a MUonE experiment~\cite{CC,Ab22} may provide a totally independent result from either the dispersive evaluations in refs.~\cite{Davier, Teubner} or the LQCD determination of ref.~\cite{BMWmu}.

We are aware of the fact that the {\it reconstruction approximants} proposed in this paper can  also be applied to many other cases; in particular to  the Euclidean representation of $a_{\mu}^{\rm HVP}$ at next-to-leading order, recently discussed in ref.~\cite{BLP}. Any observable governed by integrals of two-point functions is a possible candidate. 

 The time momentum representation proposed in ref.~\cite{BM11} for LQCD evaluations of $a_{\mu}^{\rm HVP}$, is also a particularly interesting example. It requires the determination of the Laplace-like transform:
\be
G(x_0)=\int_{\sqrt{t_0}}^\infty d\omega\   e^{-\omega{x}_0
}\ \omega^2 \frac{1}{\pi}\Imm\Pi^{\rm HVP}(\omega^2)\quad {\rm where}\quad  \omega\equiv\sqrt{t}\,,
\ee
in the full $0\le x_0 \le\infty$ range. LQCD evaluations of $G(x_0)$ with good precision are, however,  limited to a restricted $x_0$-window because of technical lattice constraints. The {\it reconstruction approximants} can then  be applied to obtain the necessary extrapolation of the $G(x_0)$ function to its full $x_0$ region, much the same way as we have done for the MUonE proposal.  We plan to discuss this in a forthcoming publication.

\acknowledgments{We wish to thank J\'{e}r\^{o}me Charles for joint work on possible  HVP approximants and  their application to LQCD evaluations of $a_{\mu}^{\rm HVP}$, which lead to this work. Discussions with Marc Knecht and Laurent Lellouch on $g_{\mu}-2$ in general, have also been very helpful. We are particularly grateful to J\'{e}r\^{o}me Charles, Marc Knecht and Santi Peris for comments and a careful reading of the manuscript.}

\appendix

\section{QED Vacuum Polarization  as a Theoretical  Laboratory}

At the one loop level, the QED spectral function of a fermion of arbitrary mass M  is given by~\footnote{Notice that this spectral function is the same as the one of the constituent quark model for a quark of mass M. All the relevant equations in this section are  in $\frac{\alpha}{\pi}$ units.} 
\begin{equation}
  \lbl{eq:ImPiexact}
  \frac{1}{\pi}\Imm\Pi_{\rm QED}(t) = \frac{1}{3}\left(1+\frac{1}{2}\frac{{\rm th}}{t}\right)\sqrt{1-\frac{{\rm th}}{t}} \, \vartheta(t-{\rm th})\,,\quad {{\rm th}} \equiv 4{\rm M}^2\,,
\end{equation}
and the corresponding  photon self-energy function  is
\begin{equation}
  \label{eq:Piexact}
  \Pi_{\rm QED}(-Q^2)=-\frac{1}{9z^2}\left[ (3-5z) -3\sqrt{z(1+z)}(1-2z)\arcsin\sqrt{z} \right]\,, \quad  z\equiv \frac{Q^2}{{\rm th}}\,.
\end{equation}
The Mellin transform in this case  is a  ratio of $\Gamma$-functions
\begin{equation}
  \label{eq:Mellin}
  \mathcal{M}_{\rm QED}(s)= \int_{{\rm th}}^\infty \frac{dt}{t} \left(\frac{t}{{\rm th}}\right)^{s-1}  \frac{1}{\pi}\Imm\Pi_{\rm QED}(t) =  \frac{\sqrt{\pi}}{4} \frac{\Gamma(3-s)}{(1-s)\Gamma(\frac{7}{2}-s)}\,,
\end{equation}
with simple poles at $s=1,3,4,\cdots\ .$
The coefficients of the Taylor expansion in the $\omega$-disc
\be\lbl{eq:tayloromega}
\Pi_\mathrm{QED}\left(-\frac{4\omega}{(1-\omega)^2}\right) \underset{|\omega|<1}{\sim}  \sum_{n=1}^\infty \Omega_{\mathrm{QED}}(n)  \; \omega^n\,, 
\ee
can then be evaluated analytically using Eq.~\rf{eq:OmegaasM}:

{\setl
\bea\lbl{eq:omegaqed}
\Omega_{\rm QED}(n) & = &  \frac{\sqrt{\pi}}{4}  \sum_{p=1}^{n} (-1)^{p}\  \frac{4^{p} \Gamma(n+p)}{\Gamma(2p)\Gamma(n+1-p)}\ \frac{\Gamma(2+p)}{p\ \Gamma(\frac{5}{2}+p)} \nn \\
 & & \nn \\ 
  & = & \left(-\frac{2}{3}\right)  \frac{16n^3-40n}{16n^4-40n^2+9}\,.
\eea}

\noi
In this example we have, therefore, all the ingredients to apply the transfer theorem analytically.

\vspace*{0.5cm}
 
\subsection{Reconstruction Approximants in the QED Example}

\vspace*{0.5cm}
\noi
Using the partial fraction expansion
\be
\frac{16n^3-40n}{16n^4-40n^2+9}  =  \frac{9}{8}\frac{1}{2n-1}+\frac{9}{8}\frac{1}{2n+1}-\frac{1}{8}\frac{1}{2n+3}-\frac{1}{8}\frac{1}{2n-3}\,,
\ee
one can show
that for a fixed  $n$, the coefficients $\Omega_{\rm QED}(n)$ can be written as an infinite series of inverse $n$-powers~\footnote{This  follows from the fact that:
\bea
&  & \frac{1}{2n-1}+\frac{9}{8}\frac{1}{2n+1}-\frac{1}{8}\frac{1}{2n+3}-\frac{1}{8}\frac{1}{2n-3}\nn \\
 & = & \frac{1}{8}\left[9\frac{1}{2n}\sum_{l=0}^{\infty} \left(-\frac{1}{2n} \right)^l + 9\frac{1}{2n}\sum_{l=0}^{\infty} \left(\frac{1}{2n} \right)^l - \frac{1}{2n}\sum_{l=0}^\infty\left( \frac{3}{2n}\right)^l - \frac{1}{2n}\sum_{l=0}^\infty\left( -\frac{3}{2n}\right)^l \right]\nn \\
 & = & \frac{1}{8}\frac{1}{2n}\left[9\sum_{l=0}^\infty \frac{(-1)^l +1}{2^l n^l}- \sum_{l=0}^\infty  \frac{3^l +(-3)^l}{2^l n^l}\right]\underset{{\rm odd~terms~sum}\ra0}{=}\frac{1}{8}\frac{1}{n}\sum_{p=0}^\infty\frac{9-3^{2p}}{2^{2p}n^{2p}}= \frac{1}{8}\sum_{p=0}^\infty\frac{(9-9^p)}{2^{2p} n^{2p+1}}\,.\nn
\eea}:
\be\lbl{eq:kseries}
\Omega_{\rm QED}(n)  =  \left(-\frac{2}{3}\right) \frac{1}{8}\sum_{p=0}^\infty\frac{(9-9^p)}{2^{2p}}\frac{1}{n^{2p+1}}\,.
\ee

\noi
As a result,  the asymptotic $\Omega_{\rm QED}^{\rm AS}(n)$ series generated  by the full  non-analytic structure of the $\Pi_{\rm QED}(-Q^2)$ function in the conformal disc is  {known}:
\be
 \Omega_{\rm QED}^{\rm AS}(n)=  -\frac{2}{3}\  \frac{1}{n}+\frac{3}{8}\ \frac{1}{n^5}+\frac{15}{16}\ \frac{1}{ n^7}   + \frac{273}{128}\frac{1}{n^9}+ \mathcal{O}(n^{-11})  \,.
\ee
Retaining only the leading $\cO(\frac{1}{n})$ term of this series corresponds to what was done in  section III.1. We can now improve on that by including successive terms from the series in Eq.~{ \rf{eq:QCDN}} up to an arbitrary number of L terms:
\begin{equation}\lbl{eq:ASL}
  \Omega_{\mathrm{QED}}^{\mathrm{AS}}(n,{\rm L}) = \sum_{l=1}^{\rm L} \frac{\mathcal{B}_l^{\mathrm{QED}}}{n^l}\;,
\end{equation}
where (for $p$ integer)
\be\lbl{eq:BQED}
   \mathcal{B}_{2p}^{\mathrm{QED}} = 0 \; \; \text{and }\; \; \mathcal{B}_{2p-1}^{\mathrm{QED}} = \left(-\frac{2}{3}\right)\frac{1}{8}\frac{9-9^{p-1}}{2^{2p-2}}\,.
\ee

Applying now the definition of the polylogarithm function in Eq.~{\rf{eq:polyL}} to the infinite $n$-sum
\begin{equation}
\sum_{n=1}^\infty  \Omega_{\mathrm{QED}}^{\mathrm{AS}}(n,{\rm L}) \, \omega^n\;,
\end{equation}
one gets the corresponding $\Omega_{\mathrm{QED}}^{\mathrm{AS}\,({\rm L})}(\omega)$-function as a linear combination of polylog functions, in this case with analytically known  coefficients,
\begin{equation}
\sum_{n=1}^\infty \Omega_{\mathrm{QED}}^{\mathrm{AS}}(n,{\rm L}) \, \omega^n = \Omega_{\mathrm{QED}}^{\mathrm{AS}\,(L)}(\omega) =\left(-\frac{2}{3}\right)\frac{1}{8} \sum_{p=1}^{\left\lfloor \frac{L+1}{2}\right\rfloor} \frac{9-9^{p-1}}{2^{2p-2}} \operatorname{Li}_{2p-1}(\omega)  \;.
\end{equation}

We conclude that, in this QED example, the {\it reconstruction approximants} to the lowest order self-energy function which follow from the transfer theorem  are: 

{\setl
\bea\lbl{eq:QEDNL}
\lefteqn{\Pi_{\rm QED}^{\rm N,L}\left(- \frac{x^2}{1-x}m_{\mu}^2\right)  =  \sum_{n=1}^{{\rm N}} \cA_{\rm QED}(n,{\rm L})  \ \left(\frac{\sqrt{1+\frac{x^2}{1-x}\frac{m_{\mu}^2}{\rm th}}-1}{\sqrt{1+\frac{x^2}{1-x}\frac{m_{\mu}^2}{\rm th}}+1}  \right)^n} \nn  \\
&  & +\sum_{p=1}^{\left\lfloor\frac{{\rm L}+1}{2}\right\rfloor} \ \cB_{2p-1}^{ \rm QED}\ {\rm Li}_{2\mathsf{p}-1}\left(\frac{\sqrt{1+\frac{x^2}{1-x}\frac{m_{\mu}^2}{\rm th}}-1}{\sqrt{1+\frac{x^2}{1-x}\frac{m_{\mu}^2}{\rm th}}+1}  \right)\,,
\eea}

\noi
where the coefficients
\be    
\cA_{\rm QED}(n,{\rm L}) \equiv \Omega_{\rm QED}(n)-\Omega_{\rm QED}^{\rm AS}(n,{\rm L})\quad {\rm and}\quad \cB_{2p-1}^{\rm QED}\,,
\ee
are all known analytically: 
$\Omega_{\rm QED}(n)$ in Eq.~\rf{eq:omegaqed},   $\Omega_{\rm QED}^{\rm AS}(n,{\rm L})$ in Eq.~\rf{eq:ASL}, and  $\cB_{2p-1}^{ \rm QED}$ in Eq.~\rf{eq:BQED}.

\subsection{Reconstruction Approximants of the QED Spectral Function}

The contribution to the imaginary part from  the first line of Eq.~\rf{eq:QEDNL} is the same as the one in the first line  in  Eq.~\rf{eq:fps1}. The contribution  from the polylog sums in the second line of Eq.~\rf{eq:QEDNL} can be obtained  
from the definition of the phase $\varphi$ in   Eq.~\rf{eq:omegas} and the polylog  definition in Eq.~\rf{eq:polyL}:
\be
\operatorname{Li}_{2p-1}\left( e^{i\varphi}\right)=\sum_{n=1}^\infty \frac{e^{in\varphi}}{n^{2p+1}}=\sum_{n=1}^\infty \frac{\cos{n\varphi}}{n^{2p+1}}+i\sum_{n=1}^\infty \frac{\sin{n\varphi}}{n^{2p+1}}\,.
\ee
The sum which generates the imaginary part  can  be expressed in terms of the Fourier expansions of the Bernoulli polynomials: $\operatorname{B}_{2p-1}\left(\frac{\varphi}{2\pi}\right)$, because of the well known relation
\begin{equation}
  \sum_{n=1}^\infty \frac{\sin(n\varphi)}{n^{2p-1}} = (-1)^{p}\frac{1}{2}\frac{(2\pi)^{2p-1}}{(2p-1)!} \operatorname{B}_{2p-1}\left(\frac{\varphi}{2\pi}\right)\;,
\end{equation}
and therefore,
\begin{equation}
  \Imm \left[\operatorname{Li}_{2p-1}(e^{i\varphi})\right] = (-1)^{p}\frac{1}{2}\frac{(2\pi)^{2p-1}}{(2p-1)!} \operatorname{B}_{2p-1}\left(\frac{\varphi}{2\pi}\right)\,.
\end{equation}
All the ingredients to obtain the QED spectral function {\it reconstruction approximants} are now there,  with the result:

{\setl
\bea\lbl{eq:spectLNQED}
\lefteqn{
\frac{1}{\pi}\Imm\Pi_{\rm QED}^{\rm N,L}(t)  = \left\{  -\frac{1}{\pi}\ \frac{ 2\sqrt{\frac{t}{\rm th}-1}}{\frac{t}{\rm th}}\  \sum_{n=1}^{{\rm N}}\cA_{\rm QED}(n,{\rm L}) \ {\rm U}_{n-1}\left(\frac{t-2{\rm th}}{t}\right)\right.} \nn \\ & + & \left.    \sum_{p=1}^{\left\lfloor \frac{L+1}{2}\right\rfloor} \mathcal{B}_{2p-1}^{\mathrm{QED}} (-1)^{p-1}\frac{(2\pi)^{2p-2}}{(2p-1)!} \operatorname{B}_{2p-1}\left(\frac{\pi - 2 \arctan\left(\sqrt{\frac{t}{\rm{th}}-1}\right)}{2\pi}\right)\right\}\theta(t-{\rm th})\,.
\eea}

\noi
Notice that $\cB_{3}^{\rm QED}=0$ due to the fact  that in QED  the coefficient $\mathsf{\cR}_{\mathsf{2},1}$ in the corresponding { singular series expansion} to Eq.~\rf{eq:sse} vanishes. In other words, there is no $\frac{m^2}{Q^2}\log\frac{Q^2}{m^2}$ term~\footnote{Incidentally, this is also the case in QCD for the contribution from  light quarks in the chiral limit because there is no intrinsic operator of dimension two in QCD.  However, as shown in refs.~\cite{GP,GMP,GM}, this is different for the heavy quarks contributions.} in the large-$Q^2$ expansion of the self-energy $\Pi_{\rm QED}(-Q^2)$.

\begin{figure}[!ht]
\begin{center}
\hspace*{-1cm}\includegraphics[width=0.70\textwidth]{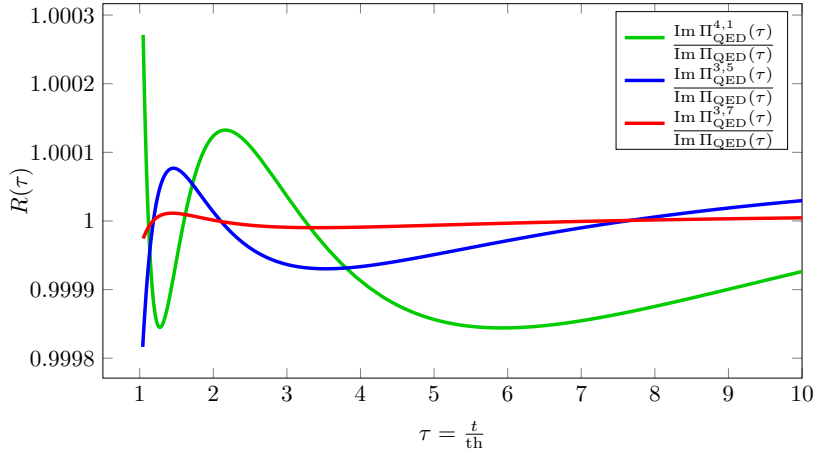} 
\caption{\lbl{fig:QEDratios} Plot of the ratios of spectral functions in Eq.~\rf{eq:ratiosspec}. The vertical scale in the figure  is very close to the identity.}
\end{center}
\end{figure}
\noi

In particular, the spectral function approximant when L=1 reduces to 

{\setl
\bea\lbl{eq:spQEDL1}
\frac{1}{\pi}\Imm\Pi_{\rm QED}^{{\rm N},{\rm L}=1}(t) & = & \left\{ -\frac{1}{\pi}\ \frac{ 2\sqrt{\frac{t}{\rm th}-1}}{\tau}\  \sum_{n=1}^{{\rm N}}\left(\Omega_{\rm QED}(n)+\frac{\alpha}{\pi}\frac{2}{3}\right) \ {\rm U}_{n-1}\left(\frac{t-2{\rm th}}{t}\right)\right. \nn \\
&  &  +  \left.  \frac{2}{\pi}\ \frac{1}{3}\  \arctan(\sqrt{\frac{t}{\rm th}-1})\right\} \theta(t -{\rm th})\,,
\eea}

\noi
which is the equivalent QED result to the one discussed in Section III.1. 

Numerically, the $\frac{1}{\pi}\Imm\Pi_{\rm QED}^{\rm N,{\rm L}=1}(t)$ approximants  reproduce the exact spectral function in Eq.~\rf{eq:ImPiexact} very fast: the shape of the $\frac{1}{\pi}\Imm\Pi_{\rm QED}^{4,1}(\tau)$ approximant for example,  already looks  identical to the one of  $\frac{1}{\pi}\Imm\Pi_{\rm QED}(\tau)$. In order to quantify the quality of the successive {\it reconstruction approximants},  we plot  in Fig~\rf{fig:QEDratios}  the ratios $\left(\tau =\frac{t}{{\rm th}} \right)$:
\be\lbl{eq:ratiosspec}
\frac{\frac{1}{\pi}\Imm\Pi_{\rm QED}^{4,1}(\tau)}{\frac{1}{\pi}\Imm\Pi_{\rm QED}(\tau)}\,,\quad  \frac{\frac{1}{\pi}\Imm\Pi_{\rm QED}^{3,5}(\tau)}{\frac{1}{\pi}\Imm\Pi_{\rm QED}(\tau)}\,,\quad  \frac{\frac{1}{\pi}\Imm\Pi_{\rm QED}^{3,7}(\tau)}{\frac{1}{\pi}\Imm\Pi_{\rm QED}(\tau)}
\ee
  respectively in green, blue and red. The figure shows that already the $\frac{1}{\pi}\Imm\Pi_{\rm QED}^{3,7}(\tau)$ approximant is practically a calculation of the QED spectral function in its full $\tau$-range. It also shows that the lower green and blue spectral approximants are not quite locally dual to the physical spectral function; their ratios  approach the identity with oscillations  which, however,  practically disappear  in the case of the red approximant. This is an illustration of what was mentioned before about local duality.

\section{Results on Combinatorial Analysis}
\subsection{Transfer Theorem}
\label{TransferTh}

Let us consider a function $f(\omega)$, defined in the the unit disc $\vert \omega\vert\le 1$,  which has the asymptotic behaviour
\be\lbl{eq:ask}
  f(\omega) \underset{\omega \rightarrow 1}{\sim} (1-\omega)^m \ln^k\left(\frac{1}{1-\omega}\right)\;,
\ee
with $m$ and $k$  positive integers. We want to know how this behaviour \emph{transfers } to the coefficients $f_n$ of its Taylor expansion:
\be
  f(\omega) \underset{|\omega|<1}{=}\sum_{n=1}^\infty f_n \omega^n\;.
\ee

First we observe that the r.h.s. in Eq.~\rf{eq:ask} can also be written in the following way ($k\geqslant1$):
\be
  (1-\omega)^m \ln^k\left(\frac{1}{1-\omega}\right)= \frac{\partial^k}{\partial \varepsilon^k} (1-\omega)^{m-\varepsilon} \bigg \vert_{\varepsilon=0}\,.
\ee
Applying  to this the inverse Mellin transform representation (the notation $<0,\infty>$ denotes the fundamental strip where the integral is analytic)
\be
  (1-\omega)^a \;\vartheta(1-\omega) = \frac{1}{2\pi i}\int\limits_{<0,\infty>} \!ds  \ \omega^{-s} \; \frac{\Gamma(a+1)\Gamma(s)}{\Gamma(a+1+s)}\,,\quad {\rm with} \quad a=m-\varepsilon\quad\annd\quad a>0\,,  
\ee
one finds that the asymptotic behaviour of Eq.~\rf{eq:ask} can also be written in the following way

\begin{equation}
  f(\omega) \underset{\omega \rightarrow 1}{\sim}\frac{\partial^k}{\partial \varepsilon^k}\ \left( \frac{1}{2\pi i} \int\limits_{<0,\infty>} \! ds \;\omega^{-s}  \; \frac{\Gamma(m-\varepsilon+1)\Gamma(s)}{\Gamma(m-\varepsilon+1+s)}\right)\Bigg \vert_{\varepsilon=0}\;,
\end{equation}
and therefore~\cite{FGD}
\begin{equation}
 f(\omega) \underset{\omega \rightarrow 1}{\sim}\frac{\partial^k}{\partial \varepsilon^k}\left( \sum_{n=0}^\infty \frac{(-1)^n}{\left(m-\varepsilon+1\right)_{-n}} \frac{\omega^n}{(1)_n}\right)\Bigg \vert_{\varepsilon=0}\;,
\end{equation}
where we have introduced the Pochhammer symbol~\cite{Slater:1966:GHF} notation: $(a)_n \equiv \frac{\Gamma(a+n)}{\Gamma(a)}$. Using the property
\begin{equation}
  \left(m-\varepsilon+1\right)_{-n} = \frac{(-1)^n}{(\varepsilon-m)_n}\;,
\end{equation}
we get
\begin{equation}
  f(\omega) \underset{\omega \rightarrow 1}{\sim} \sum_{n=0}^\infty \left[\frac{\partial^k}{\partial \varepsilon^k} \left(\varepsilon-m\right)_{n}\right] \bigg \vert_{\varepsilon=0} \frac{\omega^n}{(1)_n}\,,
\end{equation}
and the question now  is to find the asymptotic behaviour for large $n$ of the coefficients
\begin{equation}
  \frac{1}{(1)_n}\left[\frac{\partial^k}{\partial \varepsilon^k} \left(\varepsilon-m\right)_{n}\right] \bigg \vert_{\varepsilon=0}\;.
\end{equation}
The answer to that is given in Eq.~(52) of ref.\cite{Greynat:2014jsa}~\footnote{In  this reference,  the derivative corresponds to the notation $\mathcal{P}_n^{(k)}(-m)$.}: 
\begin{equation}
  \left[\frac{\partial^k}{\partial \varepsilon^k} \left(\varepsilon-m\right)_{n}\right] \bigg \vert_{\varepsilon=0} \underset{n>m}{=} (-1)^{m+k+1}\Gamma(k+1) \sum_{j=1}^{k} (-1)^j \begin{bmatrix} m+1 \\ k-j+1 \end{bmatrix}\begin{bmatrix} n-m \\ j \end{bmatrix}\;,
\end{equation}
where $\begin{bmatrix}a \\ b \end{bmatrix}$ are \emph{unsigned Stirling numbers of the first kind}~\cite{ADAMCHIK1997119}. The asymptotic behaviour of $f_n$ coefficients is, therefore,  proportional to a linear combination of the number symbols
\begin{equation}
    \frac{1}{(1)_n}\begin{bmatrix} n-m \\ j \end{bmatrix}\,,\quad {\rm for}\quad 1 \leqslant j \leqslant k\,.
\end{equation}
When $j$ is an integer, the expressions for these symbols are well-known~\cite{ADAMCHIK1997119}. They are  given in terms of harmonic numbers;  for example,
\begin{align}
  \begin{bmatrix} n-m \\ 1 \end{bmatrix} &= (-1)^m \frac{\Gamma(n-m)}{\Gamma(1)}\\
  \begin{bmatrix} n-m \\ 2 \end{bmatrix} &= (-1)^m \frac{\Gamma(n-m)}{\Gamma(2)} \operatorname{H}_{n-m-1} \\
  \begin{bmatrix} n-m \\ 3 \end{bmatrix} &= (-1)^m \frac{\Gamma(n-m)}{\Gamma(3)} \left[\operatorname{H}_{n-m-1}^{2} -\operatorname{H}_{n-m-1}^{(2)} \right]\\
  \begin{bmatrix} n-m \\ 4 \end{bmatrix} &= (-1)^m \frac{\Gamma(n-m)}{\Gamma(4)} \left[\operatorname{H}_{n-m-1}^{3} -3\operatorname{H}_{n-m-1}\operatorname{H}_{n-m-1}^{(2)} + 2\operatorname{H}_{n-m-1}^{(3)} \right]\;,
\end{align}
where
\begin{equation}
  \operatorname{H}_{N}^{(r)} = \sum_{j=1}^{N} \frac{1}{j^r}\,,\quad {\rm and}\quad  \operatorname{H}_{N}=  \operatorname{H}_{N}^{(1)}\,,
\end{equation}
are the harmonic numbers; and their
large $N$ behaviour are well-known~\cite{Gruenberg2006}: 
\begin{align}
\operatorname{H}_{N}^{(1)} &\underset{N\rightarrow \infty}{\sim} \ln N + \sum_{j=0}^\infty \frac{h_j(1)}{N^j} \\
\operatorname{H}_{N}^{(r>1)} &\underset{N\rightarrow \infty}{\sim} \sum_{j=0}^\infty \frac{h_j(r)}{N^j} \;,
\end{align}
with coefficients
\begin{align}
  &h_0(1) = \gamma_E\;,\; \; h_1(1) = \frac{1}{2} \;,\; \; h_j(1) = - \frac{\operatorname{B}_{j}}{j}\cos^2\left(\frac{\pi}{2}j\right)\;\\
  &h_0(r) = \zeta(r)\;,\; \; h_{j<r-1}(r) = 0 \;,\;\;h_{r-1}(r) = \frac{1}{r-1}\;,\;\; h_{r}(r) =\frac{1}{2} \;, \\
  &h_{j>r}(r) =-\operatorname{B}_{j}\frac{\Gamma(j+r-1)}{\Gamma(j)\Gamma(r)}\cos^2\left(\frac{\pi}{2}j\right)\,,
\end{align}
where $\operatorname{B}_{j}$ are the  Bernoulli's numbers.

We are still left with the overall factor $\frac{\Gamma(n-m)}{(1)_n}$ in front of each linear combination of harmonic numbers, which has an expansion:
\begin{equation}
  \frac{\Gamma(n-m)}{(1)_n} =\frac{\Gamma(n-m)}{\Gamma(n+1)} =  \frac{1}{(n-m)\cdots(n-1)n} = \frac{1}{n^{m+1}}\sum_{j=0}^\infty \begin{Bmatrix} m+j \\ m \end{Bmatrix}\frac{1}{n^j}\;,
\end{equation}
where $\begin{Bmatrix} a \\ b \end{Bmatrix}$ are (this time) the \emph{Stirling numbers of the second kind}~\cite{ADAMCHIK1997119}.
Putting all the expansions together, we have the following results as explicit examples of \emph{transfers}:
\begin{align}
  (1-\omega)^m \ln\left(\frac{1}{1-\omega}\right) & \longmapsto f_n \underset{n\rightarrow \infty}{\sim} \frac{(-1)^m\Gamma(m+1)}{n^{m+1}}\sum_{j=0}^\infty \begin{Bmatrix} m+j \\ m \end{Bmatrix}\frac{1}{n^j}\\
    (1-\omega)^m \ln^2\left(\frac{1}{1-\omega}\right) & \longmapsto f_n \underset{n\rightarrow \infty}{\sim} \frac{(-1)^m\Gamma(m+1)}{n^{m+1}}\sum_{j=0}^\infty \frac{1}{n^j} \bigg[ \begin{Bmatrix} m+j \\ m \end{Bmatrix}\left(\ln n +\operatorname{H}_{m}\right) \nonumber\\
    &\hspace{6cm}+ \sum_{\ell=0}^j \begin{Bmatrix} m+\ell \\ m \end{Bmatrix} h_{j-\ell}(1)\bigg]\,,
\end{align}
where the  result in the first line corresponds to Eq.~\rf{eq:largen} in the text.

There is also a general formula for any integers $m$ and $k$ which can be obtained recursively as follows:
\be\lbl{eq:sums}
  (1-\omega)^m \ln^k\left(\frac{1}{1-\omega}\right) \longmapsto f_n \underset{n\rightarrow \infty}{\sim} \frac{(-1)^{m+k+1}\Gamma(k+1)}{n^{m+1}}\sum_{j=0}^\infty \sum_{p=0}^{k-1}g_{j,p}(m,k)\frac{\ln^p n}{n^j}\;,
\ee
where
\begin{equation}
  g_{j,p}(m,k)= \sum_{b=0}^j\sum_{a=1}^{p+1} \frac{(-1)^a}{\Gamma(a)} \begin{Bmatrix} m+j-b \\ m \end{Bmatrix} \begin{bmatrix} m+1 \\ k-a+1 \end{bmatrix} w_j(b-1,p)\;,
\end{equation}
and the coefficients $w_j(M,p)$ defined by recursive relations
\begin{align}
&  w_j(M+1,M+1) = w_j(M,M) \\
&  w_j(M+1,0) = \sum_{b=0}^j \sum_{c=0}^M (-M)_c \, h_{j-b}(c+1) \, w_j(M-c,0) \\
&  w_j(M+1,p\geqslant1) = w_j(M,p-1)+\sum_{b=0}^j \sum_{c=0}^{M-p} (-M)_c \, h_{j-b}(c+1) \, w_j(M-c,p) \;.
\end{align}
We find, in particular, that the leading term in the r.h.s. of Eq.~\rf{eq:sums}  is given by the simple expression:
\begin{equation}
  f_n \underset{n\rightarrow \infty}{=} (-1)^m \Gamma(m+1) k  \; \frac{\ln^{k-1} n }{n^{m+1}} + \mathcal{O}\left(\frac{\log^{k-2} n}{n^{m+1}} \right)\,.
\end{equation}

\subsection{Relation  between the $\cB$-coefficients and the $\cR$-residues} 
\label{Connection}

Let us start with the asymptotic expansion 
\begin{equation}
 \Pi\left(-\frac{4\omega}{(1-\omega)^2}\right) \underset{\omega \rightarrow 1}{\sim}  \sum_{m\geqslant 0} \widetilde{\mathcal{R}}_{m,1} (1-\omega)^m \ln \left(\frac{1}{1-\omega}\right) \;.
\end{equation}
The application of the transfer theorem to the r.h.s.  gives the result for the large $n$ behaviour of the Taylor coefficients: 
\begin{equation}
 (1-\omega)^m \ln \left(\frac{1}{1-\omega}\right) \longmapsto f_n \underset{n \rightarrow \infty}{\sim}    \frac{(-1)^m\Gamma(m+1)}{n^{m+1}} \sum_{j=0}^\infty \begin{Bmatrix} m+j \\ m \end{Bmatrix}\frac{1}{n^j}\;,
\end{equation}
and therefore
\begin{equation}
  \lbl{eq:proffRndB1}
\Omega^\mathrm{AS}_n =  \sum_{j=0}^\infty \sum_{m \geqslant 0}  \widetilde{\mathcal{R}}_{m,1} \begin{Bmatrix} m+j \\ m \end{Bmatrix} \frac{(-1)^m\Gamma(m+1)}{n^{m+1+j}} \;.
\end{equation}

Recall next the definition of the $\mathcal{B}_l$-coefficients in the text:
\be\lbl{eq:BRs}
\Omega^\mathrm{AS}_n =  \sum_{l=0}^\infty \frac{\mathcal{B}_l}{n^l} \;.
\ee
Separating the first term in Eq.~\rf{eq:proffRndB1}, we have
\begin{align}\lbl{eq:tildes}
\Omega^\mathrm{AS}_n & = \frac{-2\mathcal{R}_{1,1}}{n} +  \sum_{j=0}^\infty \sum_{m \geqslant 1}  \widetilde{\mathcal{R}}_{m,1} \begin{Bmatrix} m+j \\ m \end{Bmatrix} \frac{(-1)^m\Gamma(m+1)}{n^{m+1+j}}\;,
\end{align}
and therefore
\begin{equation}
 \mathcal{B}_1 = -2\mathcal{R}_{1,1}\;.
\end{equation}
Performing the change of variable:  $l=m +1 +j$ in the second term of Eq.~\rf{eq:tildes} results in the expression 
\begin{multline}
 \sum_{j=0}^\infty \sum_{m \geqslant 1}  \widetilde{\mathcal{R}}_{m,1} \begin{Bmatrix} m+j \\ m \end{Bmatrix} \frac{(-1)^m\Gamma(m+1)}{n^{m+1+j}}\\
 = \sum_{l=2}^\infty \sum_{j=0}^\infty \sum_{m \geqslant 1}  \widetilde{\mathcal{R}}_{m,1} \begin{Bmatrix} m+j \\ m \end{Bmatrix} \frac{(-1)^m\Gamma(m+1)}{n^{m+1+j}} \delta(m+1+j-l)\ \vartheta(j)\;,
\end{multline}
and reorganizing the sums   with the use of the  $\delta$-function argument, we get
\begin{equation}
 \sum_{j=0}^\infty \sum_{m \geqslant 1}  \widetilde{\mathcal{R}}_{m,1} \begin{Bmatrix} m+j \\ m \end{Bmatrix} \frac{(-1)^m\Gamma(m+1)}{n^{m+1+j}} = \sum_{l=2}^\infty \sum_{m=1}^{l-1}  \widetilde{\mathcal{R}}_{m,1} \begin{Bmatrix} l-1 \\ m \end{Bmatrix}   \frac{(-1)^m\Gamma(m+1)}{n^l} 
\end{equation}
and therefore,  by identification with Eq.~\rf{eq:BRs},  the wanted results:
\begin{align}
 \mathcal{B}_1 &= -2 \mathcal{R}_{1,1} \\
 \mathcal{B}_l &=  \sum_{m=1}^{l-1}  \widetilde{\mathcal{R}}_{m,1} \begin{Bmatrix} l-1 \\ m \end{Bmatrix} (-1)^m\Gamma(m+1) \;.
\end{align}

Using the relation
\begin{equation}
  \widetilde{\mathcal{R}}_{m,1} = -2 \sum_{\mathsf{p}=2}^{\lfloor\frac{m+2}{2}\rfloor}  \binom{m-\mathsf{p}}{\mathsf{p}-2} 4^{1-\mathsf{p}} \mathcal{R}_{\mathsf{p},1}\;,
\end{equation}
we have
\begin{equation}\lbl{eq:BallL}
  \mathcal{B}_l =  -2 \sum_{m=1}^{l-1} \sum_{\mathsf{p}=2}^{\lfloor\frac{m+2}{2}\rfloor}  \begin{Bmatrix} l-1 \\ m \end{Bmatrix} (-1)^m\Gamma(m+1)\binom{m-\mathsf{p}}{\mathsf{p}-2} 4^{1-\mathsf{p}} \mathcal{R}_{\mathsf{p},1} \;,
\end{equation}
and  more explicitly, up to $l=5$:
\begin{align}
  \mathcal{B}_2 &= 0 \\
  \mathcal{B}_3 &= \mathcal{R}_{2,1} \\
  \mathcal{B}_4 &= 0 \\
  \mathcal{B}_5 &= \mathcal{R}_{2,1} + 3 \mathcal{R}_{3,1} \;,
\end{align}
 which are the terms that we have been using in the numerical applications. 

\vspace*{0.5cm}
The fact that all the $l$-even coefficients $\mathcal{B}_{l}$ vanish,  
follows from Eq.~\rf{eq:BallL}, but there is an interesting reason for that. This property is related to the fact that the vacuum polarization self-energy, as a  function of the conformal $\omega$-variable,  is invariant under the transformation $\omega \mapsto \frac{1}{\omega}$:
\begin{equation}
  \Pi\left(-\frac{4\omega}{(1-\omega)^2}\right) \longmapsto \Pi\left(-\frac{4\frac{1}{\omega}}{(1-\frac{1}{\omega})^2}\right) =   \Pi\left(-\frac{4\omega}{(1-\omega)^2}\right)\;.
\end{equation}
Formally, for its Taylor series in Eq.~\rf{eq:PiOmegaExp}, this implies
\begin{equation}
  \sum_{n=1}^\infty \Omega_n \omega^n =\sum_{n=1}^\infty \Omega_n \left(\frac{1}{\omega}\right)^n = \sum_{n=1}^\infty \Omega_{-n} \omega^n\,,
\end{equation}
and using the relation in Eq.~\rf{eq:OmegaasM} between the $\Omega_n$-coefficients and  the Mellin moments:
\begin{equation}
  \Omega_n = \sum_{p=1}^{\infty} (-1)^{p}\mathcal{M}(1-p) \frac{4^{p} \Gamma(n+p)}{\Gamma(2p)\Gamma(n+1-p)} =  \sum_{p=1}^{\infty} (-1)^{p}\mathcal{M}(1-p) 4^{p}  \binom{n+p-1}{2p-1}\;,
\end{equation}
it is easy to show  from properties of the binomial coefficient,  that
\begin{equation}
  \Omega_n = - \Omega_{-n}\,;
\end{equation}
i.e. $\Omega_n$ is an odd function of $n$. Therefore, in  the case were the Mellin transform has only single poles, which has been assumed in this paper, there follows that
\begin{equation}
  \Omega_n \underset{n\rightarrow \infty}{\sim} \Omega_n^{\rm AS} = \sum_{l=1}^\infty \frac{\mathcal{B}_l}{n^l}= - \Omega_{-n}^{\rm AS} = -  \sum_{l=1}^\infty \frac{(-1)^l\mathcal{B}_l}{n^l}\;,
\end{equation}
which  for $k$ integer,  implies $\mathcal{B}_{2k} = 0$.



\begin{thebibliography}{99}
\bibitem{BSM21}
Interesting as they may be, there is no space to quote them all here.
\end{thebibliography}
\end{document}